\def\arxivmode{1}
\ifdefined\arxivmode\documentclass[runningheads,envcountsame]{llncs}\else\documentclass[envcountsame]{llncs}\fi

\usepackage[T1]{fontenc}
\usepackage{graphicx}
\usepackage{mathtools}
\usepackage{amsmath}
\usepackage{amssymb}
\usepackage{xcolor}
\usepackage{csquotes}
\usepackage{subcaption}

	\usepackage{amsthm}
	\usepackage{thmtools,thm-restate}
	\usepackage{array}
	\newif\ifarxiv
	\ifdefined\arxivmode\arxivtrue\fi
	\ifarxiv\usepackage{algorithm}\usepackage{algpseudocode}\fi
	\makeatletter
	\spn@wtheorem{observation}{Observation}{\bfseries}{\itshape}
	\makeatother
	
	\usepackage[hidelinks]{hyperref}
	\ifarxiv\usepackage{orcidlink}\renewcommand{\orcidID}[1]{\unskip\,\orcidlink{#1}}\fi
	\usepackage[capitalise,noabbrev]{cleveref}
	\crefname{observation}{Observation}{Observations}
	
	\newenvironment{proofsketch}[1][]{\begin{proof}[Proof sketch#1]}{\end{proof}}
	
	\graphicspath{{images/}}
	
	\definecolor{ibmred}{RGB}{220, 38, 127}
	\definecolor{ibmblue}{RGB}{100, 143, 255}
	
	\newcommand{\FV}{\ifarxiv\else{} of the full version~\cite{dijk2026full}\fi}
		
\begin{document}
		
		\title{Optimizing Diameter of $r$-Gather under Connectivity Constraints from Convex to Connected}
		\titlerunning{Diameter of Convex to Connected $r$-Gather}
		
		\author{Thomas C.\ van Dijk\inst{1}\orcidID{0000-0001-6553-7317} \and Wolf Ki\ss{}ler\inst{2}\orcidID{0009-0004-5609-5739} \and Lukas Pl\"atz\inst{2}\orcidID{0009-0006-0253-0883} \and Marena Richter\inst{3}\orcidID{0009-0007-8250-266X} \and Jonathan Rollin\inst{4}\orcidID{0000-0002-6769-7098}}
		\authorrunning{T.\ van Dijk, W.\ Ki\ss{}ler, L.\ Pl\"atz, M.\ Richter, J.\ Rollin}
		\institute{TU Eindhoven, Netherlands \and
			Ruhr-Universit\"at Bochum, Germany, \email{lukas.plaetz@rub.de} \and
			Universit\"at Bonn, Germany \and
			FernUniversit\"at in Hagen, Germany}
		
		\maketitle
		
		\begin{abstract}
			The $r$-gather problem partitions the nodes of an edge-weighted graph into clusters of at least $r$ nodes each.
			The problem is NP-hard, and known approximation algorithms may return clusters whose induced subgraphs are disconnected.
			We therefore require each cluster to be \emph{connected}, or even \emph{convex} in the existential sense that \emph{some} shortest path between any two of its members stays inside the cluster.
			We add a parameterized \emph{$\alpha$-ratio-convex} variant that interpolates between the two by bounding the induced stretch by a factor~$\alpha$.
			We measure the cost of a clustering by its \emph{diameter}, the largest pairwise distance within a cluster, rather than by the radius as in the classical formulation.
			We show that the variants form a hierarchy, that they remain NP-hard in general, and that none of them admits a polynomial-time $(2-\varepsilon)$-approximation unless $\mathrm{P}=\mathrm{NP}$.
			Our main algorithmic result is a polynomial-time algorithm for convex $r$-gather on trees.
			This contrasts with connected $k$-diameter, which remains hard even on star-like connectivity graphs.
			On general weighted graphs, we give a $4$-approximation for connected $r$-gather.
			Extending the tree algorithm to tree decompositions, we obtain an exact algorithm for connected $r$-gather that is fixed-parameter tractable in the treewidth together with $d$ and~$r$. %
			Finally, we relate the $\alpha$-ratio-convex variant to $t$-spanners, which yields further approximation guarantees.
			
			\keywords{$r$-gather \and anonymity \and approximation algorithms \and convexity \and connectivity \and $t$-spanner \and treewidth \and FPT}
		\end{abstract}
		
		\section{Introduction}\label{sec:intro}
		Grouping individuals into clusters of at least $r$ members is a standard primitive for privacy. Making every member indistinguishable from at least $r-1$ others is the idea behind $k$-anonymity, which is used in data publishing, location privacy, and microaggregation~\cite{aggarwal2010achieving,sweeney2002k}. How useful such a grouping is depends on how tight its clusters are, and optimizing this tightness is the problem we study.
		
\begin{figure}[t]
			\centering
			\begin{subfigure}[b]{0.23\linewidth}
				\centering
				\includegraphics[width=\linewidth, page=1]{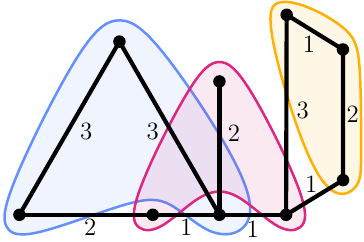}
				\caption{$d^*=3$}
				\label{fig:unrestricted_example}
			\end{subfigure}
			\hfill
			\begin{subfigure}[b]{0.23\linewidth}
				\centering
				\includegraphics[width=\linewidth, page=2]{small_example_on_constraint_r-gather.pdf}
				\caption{$d^*_\mathrm{conn}=4$}
				\label{fig:connected_example}
			\end{subfigure}
			\hfill
			\begin{subfigure}[b]{0.23\linewidth}
				\centering
				\includegraphics[width=\linewidth, page=3]{small_example_on_constraint_r-gather.pdf}
				\caption{$d^*_\mathrm{conv}=5$}
				\label{fig:convex_example}
			\end{subfigure}
			\caption{Optima of the three variants on one graph, $r=3$.}
			\label{fig:small-examples}
		\end{figure}

		In the $r$-gather problem, we are given the nodes of an edge-weighted graph with the shortest-path metric and partition them into clusters of size at least~$r$. Classically, the objective is to minimize each cluster's \emph{radius}, the largest distance to a chosen center~\cite{aggarwal2010achieving}. Since this formulation constrains only distances, its optimal solutions may have \emph{disconnected} clusters that are scattered across the graph. Such clusters are of no use in applications that must respect the topology, such as the Anonymous Routing Problem~\cite{buchin2023anonymous} that motivates this work. We therefore require each cluster to be \emph{connected} or even \emph{convex}. A cluster is convex if some shortest path between any two of its members lies inside the cluster, so that the cluster preserves the original distances (\cref{fig:small-examples}).

		Since convexity can be far more expensive than connectivity (\cref{lem:convex_unbounded_connected}), we also study a tunable relaxation, the \emph{$\alpha$-ratio-convex} variant. It requires only that between every pair of cluster members some internal path has length at most $\alpha$ times their graph distance, interpolating between convex ($\alpha = 1$) and merely connected (large~$\alpha$).
		Since these constraints concern all pairs of members, the diameter is the natural cost measure. It is symmetric, so no member is singled out as a reference point, and, like the constraints, it is defined over pairs of members. This diameter version of $r$-gather has been studied before~\cite{kumabe2021spider,zeng2017mobile}.
		
		\paragraph{Motivation.}
		These variants form a progression of increasingly topology-aware requirements, all of which arise in the Anonymous Routing Problem~\cite{buchin2023anonymous}. There, a sensitive origin or destination is replaced by another member of its cluster. In \cref{rem:routing}, we show that the three constraints correspond exactly to the three quality parameters of this scheme.
		
		Both $r$-gather and $r$-gathering are NP-hard on general graphs~\cite{aggarwal2010achieving,armon2011min} and even on weighted spider graphs~\cite{kumabe2021spider}. In \emph{$r$-gathering}, customers are assigned to open facilities, each serving at least $r$ customers, so that the largest distance between a customer and its facility is minimized~\cite{armon2011min}. Approximation algorithms achieving the optimal ratio~$2$ for $r$-gather with the radius objective~\cite{aggarwal2010achieving} and~$3$ for $r$-gathering~\cite{armon2011min} are known, and min-sum variants of $r$-gathering are solvable exactly on trees~\cite{kumabe2019tree}.
		
		Throughout, convexity is \emph{existential}. We only require that \emph{some} shortest path between two cluster members stays inside the cluster. This is what anonymous routing needs, and it is exactly what the induced distances certify. The standard notion asks this of \emph{all} shortest paths~\cite{artigas2011partitioning,pelayo2013geodesic}. The difference matters, since for every $d$ and $r\ge4$ there are planar graphs on which every $\forall$-convex $r$-gather has diameter at least~$d$, while the optimal $\exists$-convex one has diameter~$2$ (\cref{sec:notions}). From now on, we drop the quantifier.
		
		\paragraph{Our contribution.}
		We define and compare four variants of diameter $r$-gather, forming a spectrum from convex to connected, and relate them to the routing application they come from (\cref{sec:relations_rgather}, formalized in \cref{sec:app:relations}). Convex solutions can be arbitrarily larger than connected ones, connected solutions are only a constant factor larger than unrestricted ones, and all variants are NP-hard to approximate within any factor below~$2$.
		\Cref{sec:poly_trees} solves the constrained variants exactly, in $O(nr\log r)$ time on unweighted trees and in $O(n^2\log n)$ time on node-weighted trees with unit edge lengths and on edge-weighted trees (\cref{thm:conv_tree_fast}). On trees, the three constrained objectives coincide. This contrasts with the unrestricted version, which is NP-hard already on weighted trees~\cite{kumabe2021spider}, and with connected $k$-diameter clustering, which admits no $(2-\varepsilon)$-approximation already on stars~\cite{drexler2024connected}.
		For general graphs, \cref{sec:approx_alg} gives a polynomial-time $4$-approximation for connected $r$-gather. It also shows how to transfer a clustering from a $t$-spanner back to the original graph. As a consequence, a minimum spanning tree with stretch~$t$ yields a $t$-ratio-convex solution with approximation factor~$3t$.
		\Cref{sec:treewidth} lifts the tree algorithm to tree decompositions. The key step is to determine what a separator has to remember about a partial solution. We show that a set of truncated distance profiles per cluster class, together with one saturated weight, suffices. This gives an exact algorithm for connected $r$-gather in $2^{(d+2)^{O(\mathrm{tw})}}(r+1)^{O(\mathrm{tw})}n+O(n\lvert E\rvert+n^2\log n)$ time for integral edge lengths.%
		\footnote{A preliminary version of parts of Sections~\ref{sec:relations_rgather}--\ref{sec:approx_alg} was presented at EuroCG~2025~\cite{dijk2025spectrum}.}\ifarxiv\else{} The appendices and all statements numbered above~\ref{cor:tw-fpt} are found in the full version~\cite{dijk2026full}.\fi

		\paragraph{Related work and terminology.}
		The problem of $r$-gather appears in the literature under several names, including \enquote{facility location}~\cite{celik2020comparative,svitkina2010lower}, \enquote{microaggregation}~\cite{carvalho2023survey,domingo2002practical}, and \enquote{clustering with lower-bounded cluster sizes}~\cite{abu2018clustering}. Bounding the diameter and maximizing the cluster size instead gives the $s$-club problem~\cite{dondi2019covering,schafer2012parameterized}. Connectivity constraints appear in $r$-gathering with nearest-center assignment~\cite{armon2011min,kumabe2019tree} and in $k$-Anonymous Steiner Forest~\cite{haunert2021anonymization}, but to our knowledge they have not been studied for the diameter objective of $r$-gather. The convex variants appear to be new in this context.
		
		\paragraph{Comparison to previous work.}
		A related line of work is the \emph{connected $k$-center} problem~\cite{ge2008joint} and its diameter counterpart~\cite{drexler2024connected}. In these problems, a metric and a separate \emph{connectivity graph} are given, every cluster must induce a connected subgraph of the connectivity graph, and there are at most $k$ clusters. Drexler, Eube, Luo, Reineccius, R{\"o}glin, Schmidt, and Wargalla~\cite{drexler2024connected} give an $O(\log^2 k)$-approximation for general metrics and show that connected $k$-\emph{diameter} admits no $(2-\varepsilon)$-approximation even on stars.
		Two differences prevent us from applying these results. First, their metric is independent of the connectivity graph, whereas ours \emph{is} the shortest-path metric of the graph in which clusters must be connected. Only this coupling makes convexity meaningful. Second, they bound the \emph{number} of clusters, while we bound the cluster sizes \emph{from below}, and no reduction between the two is known in either direction. Still, their hardness result for stars shows that a diameter objective with a connectivity constraint is not automatically easy on tree-like graphs, in contrast to our result for trees (\cref{thm:conv_tree_fast}).
		
		Bla{\v{z}}ej, Ganian, Knop, Pokorn{\'{y}}, Schierreich, and Simonov~\cite{blavzej2023parameterized} study network microaggregation, where every cluster must admit a \emph{center} within distance~$d$ and its size is bounded from below by $\ell$ \emph{and from above by}~$u$. They solve the connected variant in $O(n^3)+d^{O(\mathrm{tw}^2)}u^{O(\mathrm{tw})}n$ time, which is XP in the treewidth and fixed-parameter tractable in $\mathrm{tw}+u+d$. The upper bound is needed as a parameter, since with a structural parameter alone the problem is W[1]-hard~\cite{blavzej2023parameterized,gaikwad2026connected}. Our problem has no upper bound, and we measure the diameter rather than the radius. This changes what a bag of the tree decomposition has to remember, as we explain in \cref{sec:treewidth}. \Cref{tab:related}\FV{} compares our results with previous work.
		
		\section[Relations between Variations of r-Gather]{Relations between Variations of $\boldsymbol{r}$-Gather}
		\label{sec:relations_rgather}
		We consider undirected graphs $G=(V, E)$ with edge weights $\ell\colon E \to \mathbb{R}^+$ and use the shortest-path distance $\mathrm{d}_G(v,w)$ as the metric on~$V$. For a cluster $V'\subseteq V$ we write $\mathrm{d}_{G[V']}(v,w)$ for the length of a shortest $v$-$w$-path in the induced graph $G[V']$, with $\mathrm{d}_{G[V']}(v,w)=\infty$ if none exists. Note that $\mathrm{d}_{G[V']}(v,w)\ge \mathrm{d}_G(v,w)$. The \emph{diameter} of a cluster is $\max_{v,w\in V'}\mathrm{d}_G(v,w)$, and we assume $|V|\geq r>1$.
		
		\begin{definition}\label{def:variants}
			An \emph{$r$-gather} of $G$ is a partition of $V$ into clusters $V_1,V_2,\ldots$, each of size at least~$r$. Its cost is the largest cluster diameter $\max_i\max_{v,w\in V_i}\mathrm{d}_G(v,w)$. The \emph{(unrestricted) $r$-gather} problem asks for an $r$-gather of minimum cost. We obtain restricted variants by additionally requiring, for every cluster $V_i$:
			\begin{itemize}
				\item \emph{connected $r$-gather}: the induced subgraph $G[V_i]$ is connected;
				\item \emph{($\exists$-)convex $r$-gather}: $\mathrm{d}_{G[V_i]}(v,w)=\mathrm{d}_G(v,w)$ for all $v,w\in V_i$, i.e., some shortest $v$-$w$-path lies inside $V_i$;
				\item \emph{$\alpha$-ratio-convex $r$-gather} ($\alpha\ge 1$): $\mathrm{d}_{G[V_i]}(v,w)\le \alpha\,\mathrm{d}_G(v,w)$ for all $v,w\in V_i$.
			\end{itemize}
		\end{definition}
		
		We denote the optimal diameters by $d^*$, $d^*_\mathrm{conn}$, $d^*_\mathrm{conv}$ and $d^*_{\alpha-\mathrm{ratio}}$, omitting $G$ and~$r$ when clear from context. The $\alpha$-ratio-convex variant bounds the worst-case \emph{stretch} of every pair by~$\alpha$, recovering convexity at $\alpha=1$ and connectivity for large~$\alpha$.

		\begin{remark}\label{rem:routing}
			The three ingredients of \cref{def:variants} correspond to the three quality parameters of the cloaking scheme that motivates them~\cite{buchin2023anonymous}. In this scheme, a user hides a query endpoint $s$ in its cluster $C(s)$, the service answers the query for a substitute $s'\in C(s)$, and the user travels from $s$ to $s'$ on their own. The size bound determines the anonymity, since $s$ is indistinguishable from $r-1$ other nodes. The diameter bound determines both the resolution of the disclosed information and the error of the answer, because $\lvert \mathrm{d}_G(s,t)-\mathrm{d}_G(s',t')\rvert\le 2d$. The structure of $G[C]$ determines the route the user has to travel. A substitute can be reached without leaving~$C$, and hence without revealing that $s$ lies near the boundary of~$C$, exactly when $G[C]$ is connected. The detour has length at most $\alpha\,\mathrm{d}_G(s,s')\le\alpha d$ exactly when $C$ is $\alpha$-ratio-convex, and it can be chosen as a shortest path of~$G$ exactly when $C$ is convex. This last point is why we look beyond connectivity. In a merely connected cluster, the detour is finite, but its length is not bounded by any function of $\mathrm{d}_G(s,s')$ (\cref{lem:separations}\,(iii)). The ratios between the four optima therefore measure how much anonymity quality has to be given up for better routing. We state the scheme formally in \cref{sec:app:relations}.
		\end{remark}
		
		Throughout, we assume that $G$ is connected. If every component of $G$ has at least $r$ nodes, no optimal solution spans two components, so each component can be handled separately. Otherwise, the connected variants are infeasible and the unrestricted optimum has infinite cost (\cref{sec:app:relations}). All proofs omitted from this section are given in \cref{sec:app:relations}. The following lemma states the basic relationships between the variants.
		
		\begin{restatable}{lemma}{lemQinequalitiesQvariations} \label{lem:inequalities_variations}
			For each $\alpha \geq 1$ and each graph, $d^* \leq d^*_\mathrm{conn} \leq d^*_{\alpha-\mathrm{ratio}} \leq d^*_\mathrm{conv}$ holds.
			Moreover, for $\alpha = 1$, we have $d^*_{\alpha-\mathrm{ratio}} = d^*_\mathrm{conv}$.
		\end{restatable}
		
		\begin{proofsketch}
			Each convex solution is $\alpha$-ratio-convex, each $\alpha$-ratio-convex solution of finite cost is connected, and each connected solution is an unrestricted one. Convexity and $\alpha$-ratio convexity coincide for $\alpha=1$.
		\end{proofsketch}
		
		We first compare unrestricted and connected $r$-gather. Both optima obey the following upper bound.
		
		\begin{restatable}{lemma}{lemQUpperQBound} \label{lem:UpperBound}
			For each connected graph with $\lvert V\rvert\ge r$, it holds that $d^*_\mathrm{conn}\leq (2r-2)\,\ell_{\max}$, where $\ell_{\max}$ is the largest edge weight.
		\end{restatable}
		\begin{proofsketch}
			In a spanning tree, a cluster of diameter larger than $(2r-2)\,\ell_{\max}$ spans a path $p_0,\dots,p_m$ with $m\ge 2r-1$ edges. Every edge $p_ip_{i+1}$ with $r-1\le i\le m-r$ splits it into two connected clusters with at least $i+1\ge r$ and $m-i\ge r$ nodes. Repeat until every cluster spans at most $2r-2$ edges.
		\end{proofsketch}

		The bound counts nodes and also holds for integral $r$ if all node weights are at least~$1$.
		
\begin{figure}[t]
			\centering
			\begin{subfigure}[b]{0.40\linewidth}
				\centering
				\includegraphics[width=\linewidth, page=1]{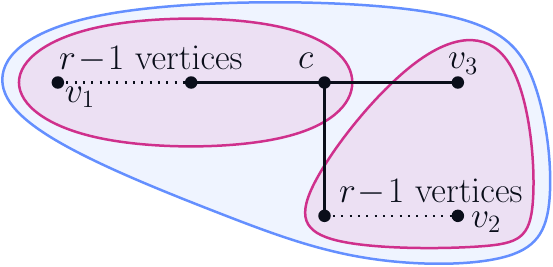}
				\caption{\cref{lem:separations}\,(i): connected (blue) vs.\ unrestricted (red) on an unweighted spider.}
				\label{fig:approximation_factor_on_spider}
			\end{subfigure}
			\hfill
			\begin{subfigure}[b]{0.53\linewidth}
				\centering
				\includegraphics[width=\linewidth]{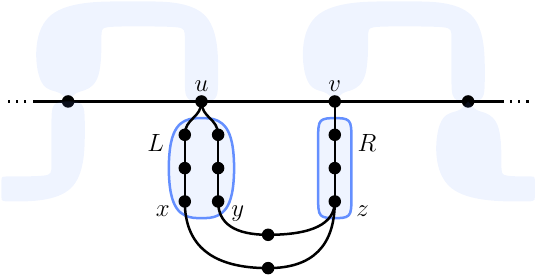}
				\caption{\cref{lem:convex_unbounded_connected}: a path of unit-weight gadgets forcing $d^*_\mathrm{conv}\ge d$ while $d^*_\mathrm{conn}$ stays in $O(r)$.}
				\label{fig:convex_unbounded_connected}
			\end{subfigure}
			\caption{Two lower-bound constructions. All edges have unit weight.}
			\label{fig:structural-examples}
		\end{figure}

		The connected and unrestricted optima differ by at most a constant factor. On unweighted spiders, the connected optimum can be almost twice the unrestricted one (\cref{fig:approximation_factor_on_spider}). Conversely, the approximation algorithms of \cref{sec:approx_alg} show that $d^*_\mathrm{conn}\le\frac72 d^*$ in general and $d^*_\mathrm{conn}\le 3d^*$ on trees. For every $\alpha>1$, the $\alpha$-ratio-convex and the connected optimum can differ by a factor larger than~$2$. \Cref{lem:separations} in \cref{sec:app:relations} collects these separations.
		
		On trees, $d^*_\mathrm{conn}=d^*_{\alpha-\mathrm{ratio}}=d^*_\mathrm{conv}$ holds for every $\alpha\ge1$, since the path between any two vertices is unique. Hence \cref{lem:separations}\,(ii) also gives $d^*_\mathrm{conv}\le 3d^*$ on trees, whereas computing the unrestricted optimum is NP-hard already on weighted trees~\cite{kumabe2021spider}. On planar graphs, the picture is completely different, since $d^*_\mathrm{conv}$ is not bounded by any function of $d^*_\mathrm{conn}$ (\cref{lem:convex_unbounded_connected} and \cref{fig:convex_unbounded_connected}).
		
		\begin{lemma} \label{lem:convex_unbounded_connected}
			For each $d > 0$ and $r\geq 12$, there is a planar unit-weight graph $G$ such that $d^*_\mathrm{conv} \geq d$ and $d^*_\mathrm{conn} \leq \lceil r/12\rceil + 8$. The same holds for $d^*_{\alpha-\mathrm{ratio}}$ whenever $1\le\alpha<\frac54$.
		\end{lemma}
		\begin{proofsketch}
			Attach to each edge $uv$ of a path of length $d$ a unit-weight gadget with eleven internal nodes. It consists of a cycle of length $10$ through $u$ and a path of three edges from $v$ to the node antipodal to~$u$. Since $r\ge12$, every cluster meets the path, and convexity then forces the two endpoints of each path edge into one cluster, so the whole graph becomes a single cluster. Conversely, every gadget node is within distance~$4$ of its path node, so merging $\lceil r/12\rceil$ consecutive path edges with their gadgets gives connected clusters of the required size and diameter at most $\lceil r/12\rceil+8$. The $\alpha$-ratio claim follows because the smallest detour ratio the argument exploits is $\frac54$.
		\end{proofsketch}

		The locality constraint of~\cite{armon2011min,kumabe2019tree}, where each node is assigned to the nearest open facility, is not comparable to ours, since it does not guarantee a constant ratio-convexity (\cref{sec:app:relations}). Finally, all of our variants are NP-hard in general.
		
		\begin{restatable}[from~\cite{zeng2017mobile}]{lemma}{lemQconvexQhardness}\label{lem:np_hardness}
			The connected, $\alpha$-ratio-convex and convex $r$-gather problems are NP-hard (for fixed $r\geq3$).
		\end{restatable}
		\begin{proofsketch}
			Zeng, Telang, Johnson, Sarkar, Gao, Arkin, and Mitchell~\cite[Thm.~4.2]{zeng2017mobile} reduce planar circuit SAT to $r$-gather using gadgets whose only clusterings of diameter~$1$ encode the Boolean values. All of these clusters are cliques of the resulting unit-disk graph and hence convex. Since they are forced, the four optima coincide on these instances, and the hardness carries over to every variant.
		\end{proofsketch}
		
		The instances produced by the reduction of \cref{lem:np_hardness} have unit weights. Their optimal diameter is~$1$ if the formula is satisfiable and at least~$2$ otherwise, which gives the following inapproximability result.
		
		\begin{corollary}[from~\cite{zeng2017mobile}]\label{cor:inapprox}
			Unless $\mathrm{P}=\mathrm{NP}$, none of the unrestricted, connected, $\alpha$-ratio-convex, or convex $r$-gather problems admits a polynomial-time $(2-\varepsilon)$-approximation for any $\varepsilon>0$ (for fixed $r\geq 3$).
		\end{corollary}

		\section{Polynomial-Time Algorithm for Trees} \label{sec:poly_trees}
		
		In this section, nodes carry non-negative weights $\mathrm{w}_v\ge0$, and the total node weight $\mathrm{W}(C)$ of a cluster replaces its size $\lvert C\rvert$, so an $r$-gather is a partition into clusters $C$ with $\mathrm{W}(C)\ge r$. We give a dynamic program that decides, for given $d$ and $r\in\mathbb{R}^+$, whether an edge- and node-weighted tree admits a connected $r$-gather in which every cluster (a subtree) has diameter at most~$d$. We call such an $r$-gather \emph{$d$-valid}. Since $d^*_{\mathrm{conn}}$ is the distance between two vertices, a binary search over the $O(n^2)$ pairwise distances yields the optimum.
		
		\begin{figure}[t]
			\centering
			\includegraphics[width=\textwidth]{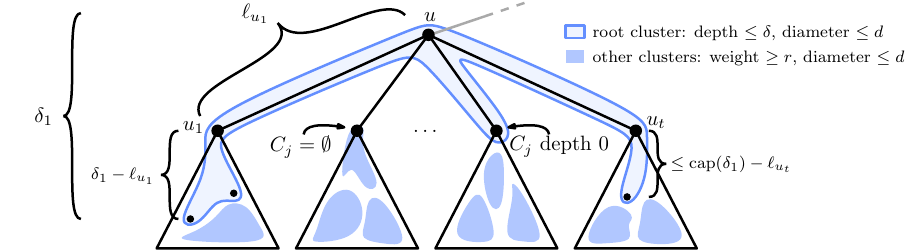}
			\caption{A $(\delta,d,r)$-valid partition, node weights not shown. Depth budgets are measured from~$u$; the subtree of $u_j$ receives the budget of its branch minus~$\ell_{u_j}$.}
			\label{fig:treeDP-example}
		\end{figure}
		
		Fix $d$ and a rooted tree $T$. Let $T_u$ be the subtree rooted at $u$, let $\mathrm{w}_v$ be the weight of node~$v$, and let $\ell_v$ be the weight of the edge from $v$ to its parent. The \emph{depth} of a rooted tree is the length of a longest root-to-leaf path.
		For $\delta \leq d$, a partition into clusters is \emph{$(\delta,d,r)$-valid} if the cluster containing the root (the \emph{root cluster}) is a subtree of depth at most $\delta$ \emph{and diameter at most~$d$}, and every other cluster is a subtree of weight at least $r$ and diameter at most $d$ (\cref{fig:treeDP-example}). Such a partition is an $r$-gather exactly when the root cluster has weight at least~$r$.

		For $\delta\le d$, let $\mathcal{R}_u(\delta)$ be the set of root clusters of the $(\delta,d,r)$-valid partitions of $T_u$. We define
		\[
		f_u(\delta)=\begin{cases}
			\max\{\mathrm{W}(C) : C\in\mathcal{R}_u(\delta)\}
			& \text{if $\delta\ge0$ and $\mathcal{R}_u(\delta)\ne\emptyset$,}\\
			0 & \text{otherwise, if $T_u$ has a $d$-valid $r$-gather,}\\
			-\infty & \text{otherwise.}
		\end{cases}
		\]
		In the first case, $u$ lies in a cluster that is still open and may be extended by the parent of~$u$. In the second case, $T_u$ is solved on its own: $u$ is not in the cluster of its parent and contributes weight~$0$ to it (the root cluster of the $d$-valid $r$-gather then obeys no depth restriction). The third case means that $T_u$ cannot be completed at all. For $\delta<0$, as it occurs in the recursion below, only the last two cases apply, so a negative argument encodes that $u$ does not join the cluster of its parent.
		Since the root cluster of a $d$-valid $r$-gather has depth at most~$d$, we obtain the following observation.
		
		\begin{observation}\label{fact:DP-rGatherExists}
			A $d$-valid $r$-gather of $T_u$ exists if and only if $f_u(d)\geq r$.
		\end{observation}

		We compute all values $f_u(\delta)$ with a dynamic program.
		For a leaf $u$, we have $f_u(\delta)=\mathrm{w}_u$ for each $\delta\geq 0$. For $\delta < 0$, we have $f_u(\delta)=-\infty$ if $\mathrm{w}_u < r$, and $f_u(\delta)=0$ otherwise.
		Otherwise, let $u_1,\ldots,u_t$ be the children of $u$.
		In the recursive step, we first compute preliminary values $\widetilde{f}_u(\delta)$ for values $\delta$ with $0\leq \delta \leq d$. \Cref{lem:dp-depths} below specifies which of these values are needed.
		Let $\ell_{u_i}$ be the length of the edge from $u$ to $u_i$. The recursion guesses a depth budget $\delta_1\in[-1,\delta]$ for one designated branch~$i$ of the root cluster, measured from~$u$, so that the child $u_i$ receives the budget $\delta_1-\ell_{u_i}$; the value $\delta_1=-1$ means that no branch joins the root cluster. Every other branch~$j$ receives the budget $\mathrm{cap}(\delta_1)-\ell_{u_j}$, where $\mathrm{cap}(\delta_1)=\min(d-\delta_1,\delta_1)$. Since $\mathrm{cap}(\delta_1)\le\delta_1\le\delta$, $\delta_1+\mathrm{cap}(\delta_1)\le d$ and $2\,\mathrm{cap}(\delta_1)\le d$, the root cluster then has depth at most~$\delta$, and any two branches together stay within diameter~$d$. Maximizing over all choices of $\delta_1$ and~$i$ gives
		\begin{equation}
			\widetilde{f}_u(\delta) = \mathrm{w}_u + \max_{-1\leq \delta_1 \leq \delta}\ \max_{i=1,\ldots,t}\left(f_{u_i}(\delta_1-\ell_{u_i}) + \sum_{j\colon j\neq i} f_{u_j}\big(\mathrm{cap}(\delta_1)-\ell_{u_j}\big)\right).\label{eq:treeDP-recursion2}
		\end{equation}
		In an optimal partition, $\delta_1$ can be taken to be the depth of the deepest branch (\cref{lem:DP-recursion}). We write $h_u(\delta_1)$ for the inner maximum over~$i$, so that $\widetilde{f}_u(\delta)=\mathrm{w}_u+\max_{-1\le\delta_1\le\delta}h_u(\delta_1)$.
		
		The cap is the only non-local part of the tree program. Once the deepest branch is chosen, it limits every other branch. The root cluster can therefore not be summarized by the weight it still needs, and depth and diameter have to be tracked together. The recursion can be evaluated efficiently for two reasons. First, the budget $\delta$ occurs only as the upper limit of the outer maximum, so a single running maximum produces the whole row for~$u$. Second, for a fixed $\delta_1$, the value $h_u(\delta_1)$ can be evaluated in $O(\deg(u))$ time. We compute the sum $S_u(\delta_1)=\sum_j f_{u_j}(\mathrm{cap}(\delta_1)-\ell_{u_j})$ once and obtain the term of branch~$i$ as $S_u(\delta_1)-f_{u_i}(\mathrm{cap}(\delta_1)-\ell_{u_i})+f_{u_i}(\delta_1-\ell_{u_i})$. Branches that contribute $-\infty$ are counted separately, so that $-\infty$ remains absorbing. \Cref{alg:treeDP}\FV{} gives pseudocode for the recursion, the grid of \cref{lem:dp-depths}, the running maximum and the correction pass described below.
		
		We now show that $\widetilde{f}_u(\delta)$ and $f_u(\delta)$ coincide if $f_u(\delta)\neq 0$.
		The cases $f_u(\delta)= 0$ and $\delta=-1$ are handled in a second step below.
		
		\begin{lemma}\label{lem:DP-recursion}
			For a non-leaf node $u$ in $T$ and $\delta\geq 0$: $f_u(\delta)\neq 0 \implies \widetilde{f}_u(\delta) = f_u(\delta)$.
		\end{lemma}
		\begin{proofsketch}
			In a maximum-weight $(\delta,d,r)$-valid partition, the two largest branch depths $\delta'_1\ge\delta'_2$ at $u$ satisfy $\delta'_1\le\delta$ and $\delta'_1+\delta'_2\le d$, so capping the non-deepest branches at $\mathrm{cap}(\delta'_1)$ loses nothing. Conversely, for any admissible $\delta_1$ and any branch $i$, the corresponding partitions of the subtrees merge into a $(\delta,d,r)$-valid partition of $T_u$.
		\end{proofsketch}
		
		In the remaining cases $f_u(\delta)=0$ and $\delta=-1$, recursion~\eqref{eq:treeDP-recursion2} returns $-\infty$ unless a root cluster of weight~$0$ exists. We correct these entries in one extra pass over the computed values and use \cref{fact:DP-rGatherExists} to decide whether $T_u$ admits a $d$-valid $r$-gather at all. \Cref{sec:app:dp} gives the details.
		
		\begin{corollary}\label{cor:tree_all_variants}
			Since the constrained objectives coincide on trees (\cref{sec:relations_rgather}), the algorithm of \cref{thm:conv_tree_fast} computes the optimal diameter of all constrained variants simultaneously, within the same time bounds.
		\end{corollary}
		
		It remains to bound the set of arguments at which the rows are evaluated. The following lemma does this without changing the computed values.
		
		\begin{lemma}\label{lem:dp-depths}
			Let $\Delta_u=\{\mathrm{d}_T(u,x) : x\in T_u\}$. Then $f_u$ is a non-decreasing step function whose breakpoints lie in $\Delta_u$, so evaluating~\eqref{eq:treeDP-recursion2} on $\Delta_u$, with $\delta_1$ restricted to $\Delta_u\cup\{-1\}$, yields the same function~$f_u$.
		\end{lemma}

		\begin{theorem}\label{thm:conv_tree_fast}
			The optimal diameter $d^*_\mathrm{conv}=d^*_\mathrm{conn}$ of a tree can be computed in $O(nr\log r)$ time on unweighted trees with unit node weights, and in $O(n^2\log n)$ time on node-weighted trees with unit edge lengths and on edge-weighted trees.
		\end{theorem}
		\begin{proofsketch}
			\sloppy One decision call costs $O(\sum_u\lvert\Delta_u\rvert\deg(u))$ time. On unweighted trees, only at most $d+2$ values of $\Delta_u\cup\{-1\}$ are relevant, and $d\le 2r-2$ holds by \cref{lem:UpperBound}, so a call costs $O(dn)$. Otherwise $\lvert\Delta_u\rvert\le\lvert T_u\rvert$ and a call costs $O(n^2)$. Binary search adds a factor of $O(\log r)$, respectively $O(\log n)$.
		\end{proofsketch}
		
		\section{Approximation Algorithms}\label{sec:approx_alg}
		Our approximation algorithm, Best-or-Fill, is a variant of Armon's Best-or-Rest algorithm for $r$-gathering~\cite{armon2011min}. The authors of~\cite{zeng2017mobile} gave a distributed sweep version of Best-or-Rest for $r$-gather. Best-or-Fill runs in two stages. In the \enquote{Best}-stage, we compute for each node $u$ a \emph{best cluster} $B_u$ that contains $u$ and has at least $r$ nodes. We process these clusters by increasing radius $\rho_u$, breaking ties arbitrarily, and mark $B_u$ if none of its nodes lies in an already marked cluster. This yields a family of disjoint marked clusters. In the \enquote{Fill}-stage, we assign every uncovered node to the nearest marked cluster and break ties in favor of the cluster marked first. Here, the distance to a cluster is the distance to its closest node, that is, to its \emph{boundary}. Measuring the distance to a center, as done in~\cite{armon2011min,zeng2017mobile}, does not preserve connectivity (\cref{sec:app:approx}).
				
		For the best cluster $B_u$, we use the simplest choice, which we call a \textsc{Type1} cluster. Let $\rho_u$ be the distance from $u$ to its $r$-th nearest node, counting $u$ itself. The cluster $B_u$ consists of all nodes at distance less than $\rho_u$ from~$u$, together with enough nodes at distance exactly $\rho_u$ to reach size~$r$. Since edge lengths are positive, every internal node of a shortest path from $u$ to a node of $B_u$ has distance less than $\rho_u$ from~$u$ and hence lies in~$B_u$. Thus $G[B_u]$ is connected. Among the cluster types we consider, only \textsc{Type1} can be computed in polynomial time for arbitrary~$r$. Three further types give better constants. A \textsc{Type2} cluster is a minimum-diameter $r$-set containing~$u$ and may be disconnected. A \textsc{Type3} cluster is its connected closure along shortest paths from~$u$. A \textsc{Type4} cluster is a minimum-diameter \emph{connected} $r$-set containing~$u$. All three types require a minimum-diameter $r$-set. We compute it by enumerating all $\binom{n}{r}$ candidate sets, which takes $n^{O(r)}$ time and is polynomial only for constant~$r$. It is at least as hard as $r$-clique detection. \Cref{sec:app:approx} defines and analyzes these types in detail.

		With \textsc{Type1} clusters, we obtain our main guarantee for general graphs.
		
		\begin{restatable}{theorem}{thmQBaF}\label{thm:approx:Type1cluster}
			Best-or-Fill with \textsc{Type1} clusters yields a connected $r$-gather solution that is a 4-approximation to the unrestricted $r$-gather problem.
		\end{restatable}
		\begin{proof}
			By construction $B_u$ is contained in the ball of radius $\rho_u$ around~$u$, so $\mathrm{diam}(B_u)\le 2\rho_u$. Let $S^*$ be an optimal unrestricted $r$-gather, of cost~$d^*$. The cluster of $S^*$ containing $u$ has at least $r$ nodes, all of them within distance $d^*$ of~$u$. Hence $\rho_u\le d^*$ and $\mathrm{diam}(B_u)\le 2\rho_u\le 2d^*$ for every $u\in V$.
			If a node $v$ lies in no marked cluster, then $B_v$ was skipped, so some $w\in B_v$ already belonged to a marked cluster. As $\mathrm{d}_G(v,w)\le\rho_v\le d^*$, the \enquote{Fill}-stage assigns $v$ to a marked cluster at distance at most $d^*$. Let $B$ be a marked cluster, and let $a,b$ be two nodes of its \emph{extended cluster}, which consists of $B$ and all nodes assigned to~$B$. Let $a',b'\in B$ be nodes of $B$ nearest to $a$ and~$b$, respectively. The triangle inequality and $\mathrm{diam}(B)\le 2d^*$ give $\mathrm{d}_G(a,b)\le \mathrm{d}_G(a,a')+\mathrm{diam}(B)+\mathrm{d}_G(b',b)\le d^*+2d^*+d^*=4d^*$, where the outer terms vanish for nodes of $B$ itself. This is the claimed approximation factor.
			
			We now show that the extended clusters are connected. Let $v$ be assigned to a marked cluster $B$ and let $x\in B$ be a closest node of~$B$. Every internal node $w$ of a shortest $v$-$x$-path lies in no marked cluster, since otherwise $w$ would be a node of a marked cluster that is closer to $v$ than~$x$. Moreover, the subpath from $w$ to $x$ is a shortest path from $w$ to~$B$, so the same tie-breaking rule assigns $w$ to~$B$ as well. Hence the whole path lies in the extended cluster of~$B$, which is therefore connected.
		\end{proof}
		
		The remaining cluster types improve the guarantees at the price of a running time of $n^{O(r)}$. They also establish \cref{lem:separations}\,(ii). All proofs are in \cref{sec:app:approx}.
		
		\begin{theorem}\label{thm:approx:Type2cluster}
			With \textsc{Type2} clusters, Best-or-Fill yields an $r$-gather that is a $3$-approximation to unrestricted $r$-gather and that is connected if $G$ is a tree. With \textsc{Type3} clusters, it yields a connected $r$-gather that is a $\frac72$-approximation to unrestricted $r$-gather. With \textsc{Type4} clusters, it yields a connected $r$-gather that is a $3$-approximation to connected $r$-gather.
		\end{theorem}
		
		For $\alpha$-ratio-convexity, we take a detour via $t$-spanners. Recall that a \emph{$t$-spanner} of $G$ is a subgraph $G_t$ in which every pairwise distance is at most $t$ times the corresponding distance in~$G$. Usually, one looks for a $t$-spanner with few edges~\cite{althofer1993sparse,narasimhan2007geometric}. The following theorem shows how a clustering computed in $G_t$ transfers back to~$G$.
		\begin{theorem}\label{thm:spanner}
			Let $G_t$ be a $t$-spanner of $G$.
			If $S$ is an $\alpha$-ratio-convex $r$-gather of $G_t$ that is a $\beta$-approximation to unrestricted $r$-gather in~$G_t$, then $S$ is a $t\alpha$-ratio-convex $r$-gather of $G$ that is a $t\beta$-approximation to unrestricted $r$-gather in~$G$.
		\end{theorem}
		\begin{proofsketch}
			Passing from $G_t$ to $G$ only adds edges, so clusters stay connected, and distances shrink by at most a factor of~$t$. For the diameters $D$ in $G$ and $D^t$ in~$G_t$, this gives $D_S\le D^t_S\le \beta D^t_{S^*}\le t\beta D_{S^*}$.
		\end{proofsketch}
		
		\begin{corollary}\label{cor:mst_spanner}
			If a minimum spanning tree of $G$ is a $t$-spanner of~$G$, then a $t$-ratio-convex $r$-gather of $G$ that is a $3t$-approximation to unrestricted $r$-gather can be computed in polynomial time.
		\end{corollary}
		
		We use a spanning tree because \cref{thm:conv_tree_fast} applies to it. The resulting guarantee is only as good as the stretch of the tree. This stretch is already $n-1$ on a unit-weight cycle, but it is small on graphs that are close to trees.
		
		\section{Bounded Treewidth}\label{sec:treewidth}
		
		We next lift the tree algorithm to graphs of bounded treewidth. Throughout this section, the edge lengths are positive integers, the node weights are non-negative integers, $G$ is connected, and we decide, for fixed $d$ and~$r$, whether $G$ admits a connected $r$-gather all of whose clusters have diameter at most~$d$. The heart of the construction is the state of the dynamic program. We describe this state and explain how it differs from the state of the radius algorithm of~\cite{blavzej2023parameterized}. The transitions (summarized in \cref{tab:tw-rules}\FV{}) and all proofs are given in \cref{sec:app:tw}.
		
		\paragraph{The adaptation.} The bags $X_t$ of a nice tree decomposition separate the processed part $V_t$ from the rest of the graph. Hence, for $x\in V_t$ and $y\notin V_t$, every shortest $x$-$y$-path passes through the bag. This gives the \emph{port identity} $\mathrm{d}_G(x,y)=\min_{b\in X_t}(\mathrm{d}_G(x,b)+\mathrm{d}_G(b,y))$ of \cref{lem:tw-port}. We call the vector of distances from a vertex to the bag vertices, truncated at~$d$, the \emph{profile} of the vertex. Profiles correspond to the \emph{center-types} of~\cite{blavzej2023parameterized}. For the radius objective, it suffices to store one profile per cluster, namely that of its center, because every constraint refers to this single anchor. A diameter bound has no such anchor, since it constrains \emph{pairs} of vertices. By the port identity, each processed vertex of a cluster reaches the unprocessed part through its own ports, and different vertices may use different ports. The state therefore stores, for each cluster class, the \emph{set} of realized profiles, reduced to its maximal elements. This antichain over $\lambda^{\mathrm{tw}+1}$ possible profiles, where $\lambda=\min(d+2,n+1)$, replaces the matrix of center-types in~\cite{blavzej2023parameterized}, which has only $d^{O(\mathrm{tw}^2)}$ possible values. The antichain is the source of the doubly exponential running time, which makes our algorithm FPT in $\mathrm{tw}+d+r$. Since $\lambda$ can be as large as $n+1$, the running time is not XP in~$\mathrm{tw}$. The size coordinate, in contrast, remains cheap. The authors of~\cite{blavzej2023parameterized} cap the accumulated size at the upper bound~$u$, since any larger size is infeasible. We cap it at~$r$, since all larger sizes are equivalent. This is \cref{lem:dp-saturation} lifted to separators (\cref{cor:tw-weight}), and it limits this coordinate to $(r+1)^{O(\mathrm{tw})}$ values instead of $n^{O(\mathrm{tw})}$.
		
		Connectivity is handled as in~\cite{blavzej2023parameterized} and in the standard \textsc{Steiner Tree} program, by refining the bag partition into connected groups.
		
		For each open part $Q$, the state records the following four entries.
		\[
		\begin{array}{c|l}
			\beta(Q) & \text{the bag vertices of the open cluster}\\
			\gamma(Q) & \text{the connectivity partition of those bag vertices}\\
			\Pi(Q) & \text{the maximal truncated distance profiles of its processed vertices}\\
			\omega(Q) & \text{its private weight, saturated at } r.
		\end{array}
		\]
		The invariant is that each open class is the processed part of one final cluster. The profiles $\Pi$ decide all future diameter constraints, and the partition $\gamma$ decides which groups can still be joined. The transitions follow from this invariant. At an introduce node, the new vertex either opens a class of its own or joins an existing class. It may join a class exactly if its profile is within distance~$d$ of every maximal profile of that class. Afterwards, all stored profiles gain a coordinate for the new vertex, which is computed by the port identity. At a forget node, all profiles are restricted to the smaller bag and re-maximized. The forgotten vertex is added to the private weight. A class that loses its last bag vertex is closed, and the state is discarded unless the saturated weight of that class equals~$r$. At a join node, we combine two states with the same bag classes. We unite and re-maximize the profile sets, merge the connectivity partitions by transitive closure, and add the private weights. The combination is discarded unless $m(p_1,p_2)\le d$ holds for all maximal profiles $p_1,p_2$ of the two sides in each class. This is the diameter test for the pairs of vertices separated by the bag. \Cref{lem:tw-transitions} shows that these rules compute exactly the reachable states.
		
		\begin{theorem}\label{thm:tw-dp}
			Let $G$ be a connected graph with $n$ vertices, $\lvert E\rvert$ edges, positive integral edge lengths, non-negative integral node weights and a given nice tree decomposition of width $\mathrm{tw}$, and let $r>1$ and $d\ge 0$ be integers. Whether $G$ admits a connected $r$-gather in which every cluster has diameter at most $d$ can be decided in $2^{\lambda^{O(\mathrm{tw})}}(r+1)^{O(\mathrm{tw})}n+O(n\lvert E\rvert+n^2\log n)$ time, where $\lambda=\min(d+2,n+1)$, and $d^*_\mathrm{conn}$ can be computed by $O(\log n)$ such calls, or by at most $O(n^2)$ calls all of which use a value $d\le d^*_\mathrm{conn}$.
		\end{theorem}
		\begin{proofsketch}
			The four transitions compute exactly the reachable states (\cref{lem:tw-transitions}), and the root bag is empty, so a partial solution there is a $d$-valid $r$-gather. \Cref{thm:tw-state} shows that the state is a sufficient statistic over the separator. The number of states per bag is bounded by the product of $2^{O(\mathrm{tw}\log\mathrm{tw})}$ partitions into classes and groups, $2^{\mathrm{tw}\lambda^{O(\mathrm{tw})}}$ antichains of profiles, and $(r+1)^{\mathrm{tw}+1}$ saturated weights. Join nodes square this number, which only doubles the exponent.
		\end{proofsketch}
		
		\begin{corollary}\label{cor:tw-fpt}
			Under the hypotheses of \cref{thm:tw-dp}, connected $r$-gather under the diameter objective is fixed-parameter tractable when parameterized by $\mathrm{tw}+d+r$. On unweighted graphs with unit node weights, it is fixed-parameter tractable when parameterized by $\mathrm{tw}+r$ alone.
		\end{corollary}

		We close this section with two remarks. First, it is open whether the profile sets can be summarized more succinctly, and hence whether the problem can be solved in polynomial time for every fixed treewidth. The natural lower-bound construction fails, because along a path inside a cluster, profiles change by at most the length of each edge. Connectivity therefore strongly restricts the profile set of an open part. Second, our algorithm only handles the connected variant. The port identity controls the \emph{length} of a shortest path across the separator but not its \emph{location}. It therefore does not certify that some shortest path stays inside the cluster, and we do not know how to decide convexity from information that is local to the bags.
		
		\section{Conclusion}\label{sec:conclusion}
		
		Motivated by anonymous routing, we introduced $r$-gather variants with connectivity and convexity constraints under the diameter objective. The variants form a hierarchy that is strict in general, and all of them are NP-hard to approximate within a factor below~$2$. On trees, however, the constrained variants coincide and can be solved in polynomial time. For general graphs, we gave a $4$-approximation for connected $r$-gather and related $\alpha$-ratio-convexity to $t$-spanners. On graphs of bounded treewidth, we gave an exact algorithm that is fixed-parameter tractable in $\mathrm{tw}+d+r$, and in $\mathrm{tw}+r$ on unweighted graphs. 
		Two questions remain open: Can the gap between the approximation factor~$4$ and the lower bound of~$2$ be closed? And is the problem solvable in polynomial time for every fixed treewidth? We do not know the answer to the second question even for $\mathrm{tw}=2$.

		\begin{credits}
			\subsubsection{\discintname}
			The authors have no competing interests to declare that are relevant to the content of this article.
		\end{credits}

		\bibliographystyle{splncs04}
		\bibliography{literature}
		
		\ifarxiv
		\clearpage
		
		\appendix

		\begin{table}[t]
			\centering\scriptsize\setlength{\tabcolsep}{3pt}
			\caption{Results on $r$-gather-type problems. Size: lower bound $r$ unless stated.}\label{tab:related}
			\begin{tabular}{@{}>{\raggedright\arraybackslash}p{0.31\linewidth}>{\raggedright\arraybackslash}p{0.16\linewidth}>{\raggedright\arraybackslash}p{0.30\linewidth}l@{}}\hline
				problem & graph class & result & ref.\\\hline
				radius $r$-gather & general metric & $2$-approx., no $(2-\varepsilon)$ for $r>6$ & \cite{aggarwal2010achieving}\\
				$r$-gathering (facilities) & general metric & $3$-approx., no $(3-\varepsilon)$ & \cite{armon2011min}\\
				diameter $r$-gather & Euclidean plane & no $1.932$-approx.\ for $r\ge3$ & \cite[Thm.~4.2]{zeng2017mobile}\\
				diameter $r$-gather, $r$-gathering & weighted spiders & strongly NP-hard, PTAS & \cite{kumabe2021spider}\\
				connected, center $\le d$, $\ell\le\lvert C\rvert\le u$ & treewidth & XP in tw, FPT in tw$+u+d$ & \cite{blavzej2023parameterized}\\
				\quad same & treewidth & W[1]-hard in tw alone & \cite{blavzej2023parameterized,gaikwad2026connected}\\
				connected $k$-diameter (no size bound) & general / stars & $O(\log^2k)$ / no $(2-\varepsilon)$ & \cite{drexler2024connected}\\\hline
				\textbf{diameter $r$-gather, all four variants} & unweighted graphs & no $(2-\varepsilon)$ for $r\ge3$ & Cor.~\ref{cor:inapprox}\\
				\textbf{connected} & general & $4$-approx. & Thm.~\ref{thm:approx:Type1cluster}\\
				\textbf{connected $=$ convex} & trees & exact, $O(n^2\log n)$ & Thm.~\ref{thm:conv_tree_fast}\\
				\textbf{connected} & treewidth & FPT in tw$+d+r$ & Thm.~\ref{thm:tw-dp}\\\hline
			\end{tabular}
		\end{table}

		\section{Notions of Convexity}\label{sec:notions}
		
		We use $\exists$-convexity, which requires that some shortest path between two cluster members stays inside the cluster. The common notion of convexity requires this for \emph{all} shortest paths. The following lemma shows that this stronger notion can be far more expensive.
		
		\begin{figure}[t]
			\centering
			\includegraphics[width=.7\linewidth]{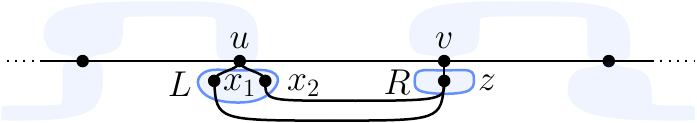}
			\caption{The construction of \cref{lem:Econvex_unbounded_connected} for $r=4$. The horizontal edges form the path, and each gadget is attached to a path edge $uv$ with $u=p_{i-1}$ and $v=p_i$.}
			\label{fig:Econvex_unbounded_connected}
		\end{figure}
		
		\begin{lemma} \label{lem:Econvex_unbounded_connected}
			For any integers $d > 0$ and $r\geq 4$, there is a planar graph $G$ such that $d^*_\mathrm{\forall-conv} \geq d$ and $d^*_\mathrm{\exists-conv} = 2$.
		\end{lemma}
		\begin{proofsketch}
			Let $p_0,\dots,p_d$ be a path with unit edges. For $i=1,\dots,d$, attach to the edge $p_{i-1}p_i$ a gadget of $r-1$ nodes (\cref{fig:Econvex_unbounded_connected}), namely nodes $x^{(i)}_1,\dots,x^{(i)}_{r-2}$, each adjacent to $p_{i-1}$ and to a node~$z^{(i)}$, which in turn is adjacent to~$p_i$. The resulting graph is planar and triangle-free, so every cluster of at least $r\ge3$ nodes has diameter at least~$2$. The gadget consists of $4$-cycles without shortcuts.
			Under $\forall$-convexity, consider gadget~$i$. Its interior has only $r-1$ nodes, so every cluster meeting it contains $p_{i-1}$ or~$p_i$. A cluster containing $x^{(i)}_j$ and $p_i$ contains both shortest paths between them and hence $p_{i-1}$. Therefore all $x^{(i)}_j$ lie in the cluster of~$p_{i-1}$. Since $r\ge4$, there are at least two of them, and the shortest path between them through $z^{(i)}$ forces $z^{(i)}$ into this cluster. The shortest path $p_{i-1},p_i,z^{(i)}$ then forces $p_i$ into it as well. Hence the whole graph collapses into one cluster of diameter at least~$d$.
			Under $\exists$-convexity, the sets $\{x^{(i)}_1,\dots,x^{(i)}_{r-2},z^{(i)},p_i\}$ for $i=1,\dots,d$, with $p_0$ added to the first one, are convex clusters of diameter~$2$.
			For this gadget, the hypothesis $r\ge4$ is necessary. For $r=3$, each gadget is a single $4$-cycle, and the triples $\{z^{(i)},p_i,x^{(i+1)}\}$ for $i=1,\dots,d-1$ are $\forall$-convex of diameter~$2$. Adding $p_0,x^{(1)}$ to the first triple and $z^{(d)},p_d$ to the last one keeps $\forall$-convexity and gives diameter~$3$. So for $r=3$ and every $d\ne2$, the $\forall$-convex optimum is at most~$3$, independently of~$d$. (For $d=2$, the first and the last triple coincide, and the optimum is~$4$.)
		\end{proofsketch}
		
		\section{Deferred Proofs for Section~\ref{sec:relations_rgather}}\label{sec:app:relations}
		
		This appendix contains the statements and proofs deferred from \cref{sec:relations_rgather}.
		
		\begin{lemma}\label{lem:separations}
			\begin{enumerate}
				\item[(i)] For every $\varepsilon>0$ there are an unweighted spider and an $r$ with $d^*_\mathrm{conn}>(2-\varepsilon)d^*$.
				\item[(ii)] For every $r$, $d^*_\mathrm{conn}\le\frac72 d^*$ on general graphs and $d^*_\mathrm{conn}\le 3d^*$ on trees.
				\item[(iii)] For every $\alpha>1$ there is a planar graph with $r=3$ and $d^*_{\alpha-\mathrm{ratio}}>2d^*_\mathrm{conn}$. In particular, the detour inside a connected cluster is not bounded by any function of the distance it replaces.
			\end{enumerate}
		\end{lemma}
		
		\begin{figure}[t]
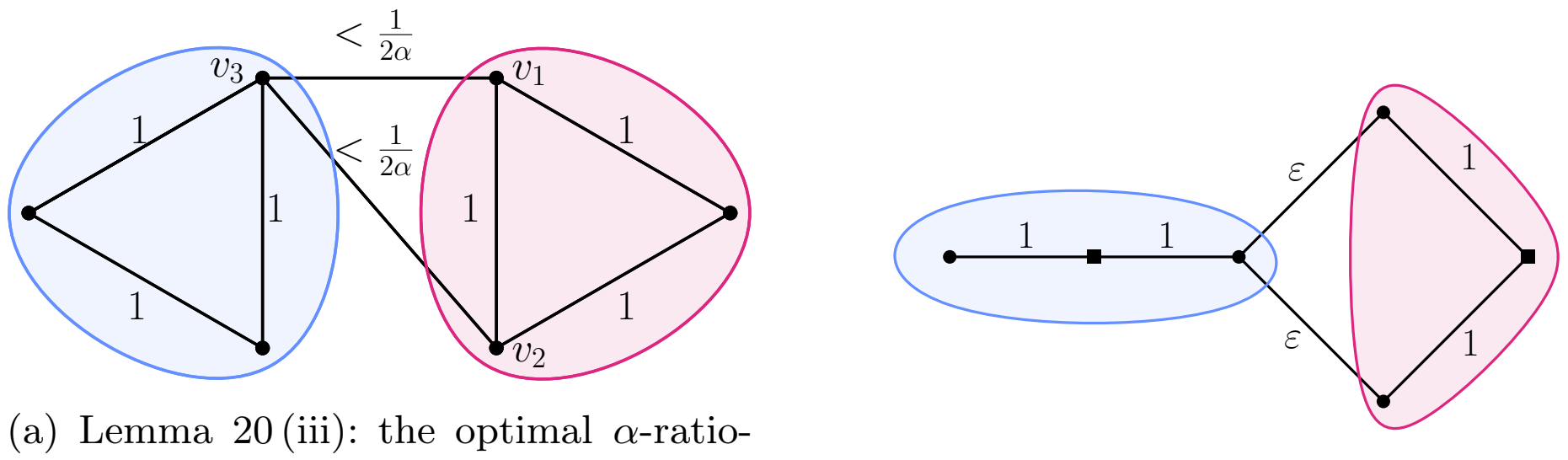

			\centering
			\begin{subfigure}[b]{0.48\linewidth}
				\centering
				\includegraphics[width=\linewidth, page=3]{rgather-all_on_pages-2.pdf}
				\caption{\cref{lem:separations}\,(iii): the optimal $\alpha$-ratio-convex solution is the whole graph, whereas the connected optimum consists of the two triangles.}
				\label{fig:alpha_convex_vs_connected}
			\end{subfigure}
			\hfill
			\begin{subfigure}[b]{0.43\linewidth}
				\centering
				\includegraphics[width=\linewidth, page=4]{rgather-all_on_pages-2.pdf}
				\caption{Assigning every node to the nearest open center gives arbitrarily bad ratio-convexity.}
				\label{fig:preferred_center_does_not_imply_constant_ratio_convexity}
			\end{subfigure}
			\caption{Further constructions separating the variants of \cref{def:variants}.}
			\label{fig:structural-examples-app}
		\end{figure}
		
		The locality constraint of~\cite{armon2011min,kumabe2019tree} assigns every node to the nearest open facility. \Cref{fig:preferred_center_does_not_imply_constant_ratio_convexity} shows that this constraint does not guarantee a constant ratio-convexity, so it is not comparable to our constraints.

		\begin{definition}\label{def:routing}
			Let $G=(V,E)$ with edge lengths $\ell$ and let $r\ge 2$. A \emph{cloaking scheme} is an $r$-gather $S=\{V_1,V_2,\dots\}$ of $G$. We write $C(v)$ for the cluster containing~$v$ and call it the \emph{anonymity set} of~$v$. A query $(s,t)$ is answered by disclosing $(C(s), C(t))$ instead of $(s,t)$ and by returning a route between a \emph{substitute} $s'\in C(s)$ and a substitute $t'\in C(t)$, together with the \emph{relay routes} from $s$ to $s'$ and from $t'$ to~$t$, which the user traverses itself. A scheme is \emph{admissible} if every relay route stays inside the corresponding anonymity set.
		\end{definition}
		
		\begin{proposition}\label{prop:routing}
			Let $S$ be an $r$-gather of $G$ of cost~$d$, and consider the scheme of \cref{def:routing}.
			\begin{enumerate}
				\item[(i)] Every node is indistinguishable from at least $r-1$ others. An adversary can localize $s$ only up to a set of diameter at most~$d$, and the reported length deviates from the true one by at most $2d$, that is, $\lvert \mathrm{d}_G(s,t)-\mathrm{d}_G(s',t')\rvert\le 2d$.
				\item[(ii)] The scheme is admissible for every choice of substitutes if and only if $S$ is connected. The relay route from $s$ to $s'$ then has length $\mathrm{d}_{G[C(s)]}(s,s')$, which is finite but in general not bounded by any function of $\mathrm{d}_G(s,s')$.
				\item[(iii)] If $S$ is $\alpha$-ratio-convex, the relay routes can be chosen of length at most $\alpha\,\mathrm{d}_G(s,s')\le\alpha\,d$. If $S$ is convex, they can be chosen as shortest paths of~$G$, so the user's detour is the unavoidable one.
			\end{enumerate}
		\end{proposition}
		\begin{proof}
			(i) The size bound is the $r$-gather constraint, and $C(s)$ has diameter at most~$d$ by assumption. For the distance, the triangle inequality gives $\mathrm{d}_G(s',t')\le \mathrm{d}_G(s',s)+\mathrm{d}_G(s,t)+\mathrm{d}_G(t,t')\le \mathrm{d}_G(s,t)+2d$ and symmetrically for the other direction.
			(ii) A relay route from $s$ to $s'$ inside $C(s)$ exists for all $s,s'\in C(s)$ if and only if $G[C(s)]$ is connected. Its length is $\mathrm{d}_{G[C(s)]}(s,s')$ by definition, and the construction of \cref{lem:separations}\,(iii) provides connected clusters in which this quantity exceeds $\mathrm{d}_G(s,s')$ by an arbitrary factor.
			(iii) Immediate from the definitions of $\alpha$-ratio-convexity and convexity.
		\end{proof}

		\begin{proof}[Proof of \cref{lem:inequalities_variations}]
			Every convex solution is $\alpha$-ratio-convex for every $\alpha\ge1$, with equality of the two notions at $\alpha=1$. Every $\alpha$-ratio-convex solution of finite cost is connected, since $\mathrm{d}_{G[V_i]}(v,w)=\infty$ for $v,w$ disconnected in $G[V_i]$. And every connected solution is an unrestricted one. The claim follows from this chain of inclusions.
			For disconnected $G$, the assumption made in \cref{sec:relations_rgather} is harmless. A cluster spanning two components has infinite diameter, so if every component has at least $r$ nodes, no optimal solution of any variant does so, and each component may be treated separately. If some component has fewer than $r$ nodes, no connected $r$-gather exists and, for $\lvert V\rvert>r$, the remaining variants have infinite optimum, so the chain holds trivially.
		\end{proof}
		
		\begin{proof}[Proof of \cref{lem:UpperBound}]
			Consider a spanning tree $T$ of $G$ and any connected $r$-gather of $T$.
			We count edges rather than weighted distances. Suppose that some $T[V']$ contains a path with at least $2r-1$ edges. Then every edge of a longest such path that has distance at least $r-1$ from both ends of the path splits $V'$ into two connected clusters of $T$ with at least $r$ nodes each. To see this, write $p_0,\dots,p_m$ for the path, where $m\ge 2r-1$. The edge $p_ip_{i+1}$ qualifies exactly if $r-1\le i\le m-r$, which is a non-empty range, and the two sides then contain at least $i+1\ge r$ and $m-i\ge r$ nodes.
			Repeatedly splitting eventually yields a connected $r$-gather of $T$ in which every cluster spans a path of at most $2r-2$ edges and hence has diameter at most $(2r-2)\,\ell_{\max}$ in $T$.
			This $r$-gather is also a connected $r$-gather of $G$, and the diameter of the clusters in $G$ is not larger than in $T$.
			The argument counts nodes. It carries over verbatim to integral $r$ and node weights at least~$1$, since each side of a split then has weight at least~$r$ as well. With general node weights, the bound fails. Consider a path $v_1,\dots,v_n$ with unit edge lengths, $\mathrm{w}_{v_1}=\mathrm{w}_{v_n}=r$ and $\mathrm{w}_{v_i}=0$ otherwise. Every cluster of weight at least $r$ contains $v_1$ or~$v_n$, so a connected $r$-gather consists of at most two subpaths and $d^*_\mathrm{conn}\ge\lfloor (n-1)/2\rfloor$ for every fixed~$r$. This is why \cref{sec:app:dp} bounds $d$ by $n$ in the node-weighted setting.
		\end{proof}
		
		\begin{proof}[Proof of \cref{lem:separations}\,(i)]
			Let $r>2/\varepsilon$ and take the spider with center~$c$, two legs $l_1,l_2$ of $r-1$ nodes and one leg $l_3$ of $r+1$ nodes. Let $v_i$ be the leaf of $l_i$ and $v_3'$ the node of $l_3$ adjacent to~$c$.
			In the unrestricted variant, the partition $\{l_1\cup\{c\},\ l_2\cup\{v_3'\},\ l_3\setminus\{v_3'\}\}$ has diameter~$r$, and no solution does better. Indeed, a cluster of at least $r$ nodes that contains $v_1$ or $v_2$ and has diameter $r-1$ must contain~$c$, but $c$ lies in only one cluster. Hence $d^*=r$.
			In the connected variant, $l_1$ and $l_2$ are too small to form clusters on their own, and their only neighbor is~$c$. Hence one cluster contains $l_1\cup l_2\cup\{c\}$ and therefore both leaves, whose distance is $2r-2$. The partition $\{l_1\cup l_2\cup\{c\},\ l_3\}$ attains this. Hence $d^*_\mathrm{conn}=2r-2$ and
			\begin{equation*}
				\frac{d^*_\mathrm{conn}}{d^*}=\frac{2r-2}{r}=2-\frac{2}{r}>2-\varepsilon.\qedhere
			\end{equation*}
		\end{proof}

		\begin{proof}[Proof of \cref{lem:convex_unbounded_connected}]
			Let $G$ be a graph obtained from a path $P$ of length $d$ by attaching a gadget to each edge.
			The gadget for the edge $uv$ consists of a $10$-cycle containing $u$ and a $4$-node path connecting $v$ and the node $z$ antipodal to $u$ in that cycle. All edges have unit weight.
			Every cluster that meets the interior of a gadget contains $u$ or~$v$. The interior has only eleven nodes, so a cluster of at least $r\ge12$ nodes must leave it, and $u$ and $v$ are the only exits.
			We name the cycle $u=c_0,c_1,\dots,c_9$ so that $z=c_5$, and we call $q_1,q_2$ the two interior nodes of the path $v,q_2,q_1,z$.
			The gadget thus contributes the $11$ internal nodes $c_1,\dots,c_9,q_1,q_2$.
			Note that the edge $uv$ of $P$ acts as a shortcut, since $\mathrm{d}_G(u,z)=\min\{5,\,1+3\}=4$, while $\mathrm{d}_G(u,c_i)=\min\{i,10-i\}$ for $i\neq 5$.
			Consider a convex $r$-gather of $G$ and any edge $uv$ in $P$.
			We show that $u$ and $v$ belong to the same cluster. Applied to every edge of~$P$, this implies that a single cluster contains the entire graph.
			Consider the gadget attached to $uv$.
			Let $L=\{c_1,c_2,c_3,c_7,c_8,c_9\}$ be the gadget nodes at distance at most $3$ from $u$ and let $R=\{q_2,q_1,z\}$ be those at distance at most $3$ from $v$ (\cref{fig:convex_unbounded_connected}).
			For each $w\in L$, the unique shortest $v$-$w$-path contains~$u$. This path has length $1+\mathrm{d}_G(u,w)\le 4$, whereas every $v$-$w$-path avoiding $u$ has to enter the cycle through $z$ and hence has length at least $3+\mathrm{d}_G(z,w)\ge 5$.
			The cluster of $w$ contains $u$ or~$v$. If it contains~$v$, convexity forces $u$ into it as well. In both cases, $w$ lies in the cluster of~$u$.
			Symmetrically, for each $w\in R$, the unique shortest $u$-$w$-path contains~$v$. It has length $1+\mathrm{d}_G(v,w)\le 4$, whereas every $u$-$w$-path avoiding $v$ runs along the cycle to $z$ and has length at least $5+\mathrm{d}_G(z,w)\ge 5$. As before, every $w\in R$ lies in the cluster of~$v$.
			Now consider the nodes $x=c_3$ and $y=c_7$, which both have distance~$3$ from~$u$.
			Their distance $\mathrm{d}_G(x,y)=4$ is realized only by the path $c_3,c_4,z,c_6,c_7$ through~$z$, since the path through $u$ has length~$6$. By convexity, $x$, $y$, and $z$ therefore lie in the same cluster.
			As $x\in L$ and $z\in R$, it follows that $u$ and $v$ are in the same cluster.
			Consequently, the entire graph lies in a single cluster of diameter at least~$d$, so $d^*_\mathrm{conv}\geq d$.
			For the upper bound on $d^*_\mathrm{conn}$, note first that every gadget node has distance at most~$4$ from the node~$u$ of $P$ to which its gadget is attached. Indeed, $\mathrm{d}_G(u,c_i)\le 4$ for all $i$, $\mathrm{d}_G(u,q_1)=3$ and $\mathrm{d}_G(u,q_2)=2$.
			(In particular, a single gadget has diameter~$5$, which is attained exactly by the four pairs $(c_1,c_6)$, $(c_2,c_7)$, $(c_3,c_8)$, and $(c_4,c_9)$.)
			Set $k=\lceil r/12\rceil$. Since the lemma only asks for \emph{some} graph with $d^*_\mathrm{conv}\ge d$, we may take the length of $P$ to be a multiple of $k$ (rounding it up only increases $d^*_\mathrm{conv}$).
			Starting from one end of~$P$, we repeatedly group $k$ consecutive path edges together with their gadgets into one block. By the choice of the length of~$P$, no edges are left over.
			Consecutive blocks share a path node, which we assign to the block on its left. This makes the resulting clusters disjoint. Every cluster receives $k$ of the $k+1$ path nodes of its block, except for the first cluster, which receives all $k+1$ of them. Each cluster is connected and contains at least $11k$ internal gadget nodes together with $k$ path nodes, that is, at least $12k\ge r$ nodes.
			Any two of its nodes $a,b$ lie in gadgets attached to path nodes $p_a,p_b$ of its block, so
			$\mathrm{d}_G(a,b)\le \mathrm{d}_G(a,p_a)+\mathrm{d}_G(p_a,p_b)+\mathrm{d}_G(p_b,b)\le 4+k+4$.
			Hence $d^*_\mathrm{conn}\leq \lceil r/12\rceil+8$, which is at most $r+6$ for every $r\ge 12$.
		\end{proof}
		
		\begin{proof}[Proof of the $\alpha$-ratio claim in \cref{lem:convex_unbounded_connected}]
			We reuse the graph and the notation of the previous proof and check that each of its three forcing steps survives under $\alpha$-ratio-convexity for $\alpha<\frac54$.
			A pair $p,q$ in a common cluster $C$ forces a node $m$ into $C$ whenever every $p$-$q$-path avoiding $m$ is longer than $\alpha\,\mathrm{d}_G(p,q)$.
			\begin{itemize}
				\item $w\in L$ and $v$ force $u$: here $\mathrm{d}_G(v,w)=1+\mathrm{d}_G(u,w)$ and the shortest $u$-avoiding $v$-$w$-path has length $3+\mathrm{d}_G(z,w)$. The ratio of the two is smallest for $w=c_3$, where it equals $\frac{5}{4}$. For $w=c_2$ it is $\frac{6}{3}=2$, and for $w=c_1$ it is $\frac{7}{2}$.
				\item $w\in R$ and $u$ force $v$: the smallest ratio here is attained for $w=z$, where $\mathrm{d}_G(u,z)=4$ and every $v$-avoiding $u$-$z$-path has length at least $5$, again a ratio of $\frac54$.
				\item $x=c_3$ and $y=c_7$ force $z$: $\mathrm{d}_G(x,y)=4$ and the only $z$-avoiding $x$-$y$-path has length $6$, a ratio of $\frac32$.
			\end{itemize}
			Since $\alpha<\frac54$ is strictly below all of these ratios, the argument of \cref{lem:convex_unbounded_connected} applies verbatim and yields $d^*_{\alpha-\mathrm{ratio}}\ge d$. The connected upper bound is unaffected.
		\end{proof}
		
		\begin{proof}[Proof of \cref{lem:separations}\,(ii)]
			With \textsc{Type3} clusters, Best-or-Fill returns a connected $r$-gather of cost at most $\frac72 d^*$ (\cref{thm:approx:Type2cluster}), so $d^*_\mathrm{conn}\le\frac72 d^*$. On trees, the same holds with \textsc{Type2} clusters and the factor~$3$.
		\end{proof}
		
		\begin{proof}[Proof of \cref{lem:separations}\,(iii)]
			Take two unit-weight triangles, fix $v_1,v_2$ in the first and $v_3$ in the second, and join them by edges $v_1v_3$ and $v_2v_3$ of length $\varepsilon<(2\alpha)^{-1}$ (\cref{fig:alpha_convex_vs_connected}). With $r=3$ the two triangles form a connected $r$-gather of diameter~$1$, so $d^*_\mathrm{conn}=1$.
			No $\alpha$-ratio-convex solution splits the graph. If $v_1,v_2$ share a cluster~$C$, then $\mathrm{d}_{G[C]}(v_1,v_2)\ge 1>2\alpha\varepsilon=\alpha\,\mathrm{d}_G(v_1,v_2)$ unless $v_3\in C$. Once $v_1,v_2,v_3$ lie in~$C$, the three remaining nodes cannot form a cluster of their own. Since $r=3$, they would have to form a single cluster, but $G-C$ is disconnected. Hence $C=V$. Now suppose that $v_1\in C_1$ and $v_2\in C_2$ with $C_1\neq C_2$. Since $\lvert V\rvert=2r$, these are the only two clusters. The node $v_3$ joins one of them, say~$C_1$. The component of $v_2$ in $G-\{v_1,v_3\}$ has only two nodes, so $C_2$ would have to extend beyond it, which violates $\alpha$-ratio-convexity.
			Hence the only solution is $C=V$, of diameter $2+\varepsilon$, and $d^*_{\alpha-\mathrm{ratio}}=2+\varepsilon>2d^*_\mathrm{conn}$.
			In the connected cluster formed by the first triangle, the detour between $v_1$ and $v_2$ has length~$1$, whereas $\mathrm{d}_G(v_1,v_2)=2\varepsilon$. Scaling all edge lengths by $x/(2\varepsilon)$ turns this into a detour of length $x/(2\varepsilon)$ for a pair at distance~$x$. Since $\varepsilon$ can be arbitrarily small, the detour is not bounded by any function of~$x$.
		\end{proof}

		\begin{proof}[Proof of \cref{lem:np_hardness}]
			Zeng, Telang, Johnson, Sarkar, Gao, Arkin, and Mitchell~\cite[Thm.~4.2]{zeng2017mobile} reduce planar circuit SAT to $r$-gather in the \emph{Euclidean} plane, where their gap is $1$ versus $\sqrt{2+\sqrt3}\approx 1.932$. We pass from their point set to a graph, which both sharpens the gap to $2$ and makes the yes-instance clusters convex. Let $G$ be the unit-disk graph on their point set, with an edge of weight~$1$ between two points at Euclidean distance at most~$1$. This graph can be built combinatorially in polynomial time: blow up every group of co-located points into a clique and join the groups that are adjacent in a gadget completely. If the formula is satisfiable, their clusters have Euclidean diameter~$1$, hence are cliques of~$G$. A clique is convex and has diameter~$1$ in $\mathrm{d}_G$, so $G$ admits a convex $r$-gather of diameter~$1$, which is feasible for every variant. If the formula is unsatisfiable, every $r$-gather contains a cluster with two points at Euclidean distance greater than~$1$. Those two are non-adjacent in $G$, so that cluster has diameter at least~$2$. Hence all four optima equal~$1$ in the yes-case and are at least~$2$ in the no-case, and each variant is NP-hard.
			We use only that their gadgets admit a partition into sets of Euclidean diameter~$1$ exactly when the circuit is satisfiable. The internal structure of the gadgets plays no further role here.
		\end{proof}
		
		\begin{proof}[Proof of \cref{cor:inapprox}]
			In the unit-weight instances of \cref{lem:np_hardness}, a satisfying assignment yields a convex $r$-gather of diameter $1$ which is valid for every variant. An unsatisfiable instance admits no clustering of diameter below $2$. A polynomial-time approximation with a factor below $2$ would distinguish the two cases and thus decide planar circuit SAT.
		\end{proof}
		
		\section{Deferred Material for Section~\ref{sec:poly_trees}}\label{sec:app:dp}

		This appendix gives the remaining details of \cref{sec:poly_trees}, namely the treatment of the corner cases of the recursion, the running time analysis, and the full proof of \cref{lem:DP-recursion}.
		
		The corner cases are the values $f_u(\delta)=0$ and $\delta=-1$, at which the recursion returns $-\infty$. They are corrected in one pass over the finished row. If $\widetilde f_u(d)\ge r$, then $T_u$ admits a $d$-valid $r$-gather by \cref{fact:DP-rGatherExists,lem:DP-recursion}, so we set $f_u(\delta)=0$ wherever $\widetilde f_u(\delta)=-\infty$, keep all other entries, and set $f_u(-1)=0$. Otherwise we keep the whole row and set $f_u(-1)=-\infty$. The only subtlety is that a $(\delta,d,r)$-valid partition whose root cluster has weight~$0$, possible when node weights vanish, already yields $\widetilde f_u(\delta)=0$ and is left untouched by the pass.
		\begin{algorithm}[t]
			\caption{Deciding whether a rooted tree $T$ admits a $d$-valid $r$-gather.}\label{alg:treeDP}
			\begin{algorithmic}[1]
				\Require $\mathrm{w}\ge0$, $\ell>0$, $d\ge0$, $r>0$; grid $\Delta_u^{(d)}=\{\delta\in\Delta_u:\delta\le d\}\cup\{-1\}$ (\cref{lem:dp-depths}). A lookup $f_c(y)$ returns the entry at the largest grid point $\le y$, and $f_c(-1)$ for $y<0$.
				\For{each node $u$ in post-order}
					\If{$u$ is a leaf} $f_u(0)\gets \mathrm{w}_u$; $f_u(-1)\gets 0$ if $\mathrm{w}_u\ge r$, else $-\infty$
					\Else
						\State $\mathit{best}\gets-\infty$; \textbf{for} $\delta_1\in\Delta_u^{(d)}$ in increasing order \textbf{do} $\mathit{best}\gets\max(\mathit{best},h_u(\delta_1))$; \textbf{if} $\delta_1\ge0$ \textbf{then} $\widetilde f_u(\delta_1)\gets\mathrm{w}_u+\mathit{best}$
						\If{$\widetilde f_u(d)\ge r$} $f_u\gets\widetilde f_u$ with every $-\infty$ replaced by $0$; $f_u(-1)\gets0$
						\Else{} $f_u\gets\widetilde f_u$; $f_u(-1)\gets-\infty$
						\EndIf
					\EndIf
				\EndFor
				\State \Return whether $f_\rho(d)\ge r$ \Comment{\cref{fact:DP-rGatherExists}}
			\end{algorithmic}
		\end{algorithm}

		\begin{proof}[Proof of \cref{lem:dp-depths}]
			A root cluster of $T_u$ is a subtree containing $u$, so its depth is the distance from $u$ to one of its nodes and lies in $\Delta_u$. Between two consecutive elements of $\Delta_u$ no additional root cluster becomes available, and $f_u$ is non-decreasing because a $(\delta,d,r)$-valid partition is $(\delta',d,r)$-valid for $\delta'\ge\delta$. For an optimal partition, the maximizing $\delta_1$ is the depth of a branch of its root cluster, measured from~$u$, and hence lies in $\Delta_u\cup\{-1\}$, while $\mathrm{w}_u+h_u(\delta_1)\le f_u(\delta)$ holds for every admissible $\delta_1$. A parent querying a non-negative argument outside $\Delta_u$ rounds down to the largest element not exceeding it. Negative arguments are answered by the convention for negative depths.
		\end{proof}
		
		For the running time of \cref{thm:conv_tree_fast}, fix $d$ and let $\Delta_u^{(d)}=\{\delta\in\Delta_u:\delta\le d\}\cup\{-1\}$ be the grid used at~$u$ (\cref{lem:dp-depths}). Evaluating $h_u(\delta_1)$ for one $\delta_1$ costs $O(\deg(u))$ by sharing $S_u(\delta_1)$ across the branches, and the running maximum over $\delta_1$ adds $O(\lvert\Delta_u^{(d)}\rvert)$, so node $u$ costs $O(\lvert\Delta_u^{(d)}\rvert\deg(u))$ and one decision call costs $O(\sum_u\lvert\Delta_u^{(d)}\rvert\deg(u))$. Lookups in the tables of the children do not require binary search. As $\delta_1$ runs through $\Delta_u^{(d)}$ in increasing order, $\mathrm{cap}(\delta_1)$ is unimodal. Hence the queries to a fixed child form three monotone runs, and each run is answered by a two-pointer scan over the sorted grid of the child. Every grid is scanned only by the parent of its node, so the total cost stays $O(\sum_u\lvert\Delta_u^{(d)}\rvert\deg(u))$. The sorted grids themselves are obtained bottom-up by merging the children's shifted grids.
		
		On an unweighted tree $\lvert\Delta_u^{(d)}\rvert\le d+2$, so one call costs $O(dn)$, and $d\le 2r-2$ by \cref{lem:UpperBound}, so binary search over $\{1,\dots,2r-2\}$ gives $O(nr\log r)$. With node weights and unit edge lengths, $d\le n$ gives $O(n^2)$ per call and $O(n^2\log n)$ overall. In the edge-weighted setting $\lvert\Delta_u^{(d)}\rvert\le\lvert T_u\rvert$ and $\sum_u\lvert\Delta_u^{(d)}\rvert\deg(u)=O(n^2)$, while the optimum is one of the $O(n^2)$ pairwise distances, which are produced and sorted once in $O(n^2\log n)$ time. Binary search then gives $O(n^2\log n)$ as well.
		
		We now give the full proof of \cref{lem:DP-recursion}.
		
		\begin{proof}[Proof of \cref{lem:DP-recursion}]
			Assume first that a $(\delta,d,r)$-valid partition of $T_u$ exists, and fix one whose root cluster $C$ has maximum weight $f_u(\delta)$. Put $C_j=C\cap V(T_{u_j})$ and let $\delta'_1\ge\delta'_2$ be the two largest depths of the extended clusters $C_j+u$, with depth $-1$ for $C_j=\emptyset$. Let $i$ be a branch attaining $\delta'_1$. Since $C$ has depth at most $\delta$ and diameter at most~$d$, we have $-1\le\delta'_2\le\delta'_1\le\delta$ and $\delta'_1+\delta'_2\le d$, also for negative values. Maximality of $C$ makes each $C_j$ heaviest under its depth bound, and for $j\neq i$ that bound may be raised to $\delta'_2-\ell_{u_j}$ without loss, so $\mathrm{W}(C_i)=f_{u_i}(\delta'_1-\ell_{u_i})$ and $\mathrm{W}(C_j)=f_{u_j}(\delta'_2-\ell_{u_j})$.
			This shows
			\begin{align*}
				f_u(\delta) &= \mathrm{W}(C) = \mathrm{w}_u + f_{u_i}(\delta'_1-\ell_{u_i}) + \sum\limits_{j\colon j\neq i} f_{u_j}(\delta'_2-\ell_{u_j})\\
				&\leq \mathrm{w}_u + \max\limits_{\substack{\delta'_1,\delta'_2\colon\\-1\leq \delta'_2 \leq \delta'_1 \leq \delta,\\ \delta'_1+\delta'_2\leq d}} \ \max_{i} \Big(f_{u_i}(\delta'_1-\ell_{u_i}) + \sum_{j\neq i} f_{u_j}(\delta'_2-\ell_{u_j})\Big) \\
				&\leq \mathrm{w}_u + \max\limits_{\substack{-1\leq \delta'_1 \leq \delta, \\ i}} \Big(f_{u_i}(\delta'_1-\ell_{u_i}) + \sum_{j\neq i} f_{u_j}(\mathrm{cap}(\delta'_1)-\ell_{u_j})\Big) = \widetilde{f}_u(\delta),
			\end{align*}
			where the last step uses $\delta'_2\le\mathrm{cap}(\delta'_1)$ and the monotonicity of each $f_{u_j}$. The last bracket is the inner maximum of~\eqref{eq:treeDP-recursion2}, in which every branch is evaluated at the cap, except for branch~$i$, which is evaluated at~$\delta'_1$.
			Conversely, consider \emph{any} $\delta_1$ with $-1\le\delta_1\le\delta$ and any branch $i$, and set $\delta_2:=\mathrm{cap}(\delta_1)$, so that $\delta_2\le\delta_1$ and $\delta_1+\delta_2\le d$.
			We show $f_u(\delta)\ge \mathrm{w}_u + f_{u_i}(\delta_1-\ell_{u_i}) + \sum_{j\neq i} f_{u_j}(\delta_2-\ell_{u_j})$, which is what \cref{lem:dp-depths} needs. Taking the maximum over $\delta_1$ and $i$ then gives $f_u(\delta)\ge\widetilde f_u(\delta)$.
			By the assumption of this case, there is a $(\delta,d,r)$-valid partition of $T_u$.
			Restricting this partition to a subtree rooted at a child of~$u$ gives either a $(\delta',d,r)$-valid partition of this subtree for some $0 \leq \delta' < \delta$ or a $d$-valid $r$-gather.
			If one of the values $f_{u_i}(\delta_1-\ell_{u_i})$, $f_{u_j}(\delta_2-\ell_{u_j})$ equals $-\infty$, the asserted inequality is trivial, because $f_u(\delta)>-\infty$ by the assumption of this case. So assume all of them are finite. (This case distinction is needed for arbitrary admissible $\delta_1$. For example, if $u$ has a single leaf child~$v$ with $\mathrm{w}_u=3$, $\mathrm{w}_v=1$, $\ell_v=1$, $r=3$, and $d=\delta=5$, then the choice $\delta_1=-1$ gives $f_v(\delta_1-\ell_v)=-\infty$, although $f_u(\delta)=4$.)
			Hence, there is a $(\delta_1-\ell_{u_i},d,r)$-valid partition of $T_{u_i}$ with root cluster of weight $f_{u_i}(\delta_1-\ell_{u_i})$ or a $d$-valid $r$-gather of $T_{u_i}$.
			For each $j\neq i$ there is a $(\delta_2-\ell_{u_j},d,r)$-valid partition of $T_{u_j}$ with root cluster of weight $f_{u_j}(\delta_2-\ell_{u_j})$ or a $d$-valid $r$-gather of $T_{u_j}$.
			Merging the root clusters of those children whose value is realized by a $(\cdot,d,r)$-valid partition, rather than by a $d$-valid $r$-gather, and adding $u$ to it gives a $(\delta,d,r)$-valid partition of~$T_u$. Its depth is at most $\max(\delta_1,0)\le\delta$, and its diameter is at most $\max\{\delta_1+\delta_2,\,d\}\le d$, using $\delta_1+\delta_2\le d$ and the fact that each merged sub-root-cluster has diameter at most~$d$. (The criterion is that the value is realized by a partition, not that the value is positive. With zero node weights, a child may contribute weight~$0$ through a genuine partition, and it must then be merged.)
			This shows that, for every admissible $\delta_1$ and every branch $i$,
			\begin{align*}
				f_u(\delta) &\geq \mathrm{w}_u + f_{u_i}(\delta_1-\ell_{u_i}) + \sum\limits_{j\colon j\neq i} f_{u_j}(\delta_2-\ell_{u_j})\\
				&\phantom{\geq}\ \text{and hence, choosing $\delta_1$ and $i$ to attain the maximum,}\\
				f_u(\delta) &\geq \mathrm{w}_u + \max\limits_{-1\leq \delta_1 \leq \delta}\ \max_{i} \Big(f_{u_i}(\delta_1-\ell_{u_i}) + \sum_{j\neq i} f_{u_j}(\mathrm{cap}(\delta_1)-\ell_{u_j})\Big) = \widetilde{f}_u(\delta).
			\end{align*}
			Now consider the case that there is neither a $d$-valid $r$-gather nor a $(\delta,d,r)$-valid partition of $T_u$ (so $f_u(\delta)=-\infty$).
			The same arguments as above show that $\widetilde{f}_u(\delta)=-\infty$ in this case.
			Indeed, if $\widetilde{f}_u(\delta)\neq -\infty$, then $\widetilde{f}_u(\delta)\ge 0$ and there are values $\delta_1$, $\delta_2$ with $-1\leq \delta_2 \leq \delta_1 \leq \delta$, $\delta_1+\delta_2\leq d$, and $f_{u_i}(\delta_1-\ell_{u_i}) + \sum_{j\colon j\neq i} f_{u_j}(\mathrm{cap}(\delta_1)-\ell_{u_j}) \geq 0$.
			We then find a $(\delta,d,r)$-valid partition of $T_u$ by merging suitable partitions of the subtrees rooted at $u$'s children.
			This shows that $\widetilde{f}_u(\delta) = f_u(\delta)$ whenever $f_u(\delta)\neq 0$.
		\end{proof}

		\section{Deferred Material for Section~\ref{sec:approx_alg}}\label{sec:app:approx}
		
		This appendix contains the material deferred from \cref{sec:approx_alg}. \Cref{fig:con-r-gather} shows why the Fill-stage measures distances to the cluster boundary. The appendix further defines the remaining cluster types, proves \cref{thm:approx:Type2cluster,lem:diameter_increas_conn,cor:mst_spanner}, and proves the $t$-spanner transfer.
		
		\begin{figure}[t]
			\centering
			\begin{subfigure}[b]{0.50\linewidth}
				\centering
				\includegraphics[width=\linewidth, page=1]{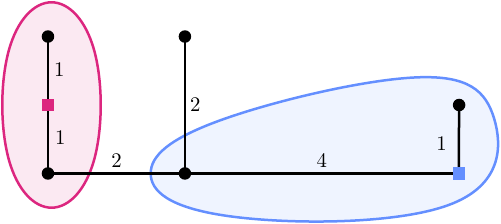}
				\caption{The Fill-stage assignment rule. Best-or-Rest would assign the remaining node to the red cluster on the left. Best-or-Fill measures the distance to the cluster boundary and keeps clusters connected.}
				\label{fig:con-r-gather}
			\end{subfigure}
			\hspace{2em}
			\begin{subfigure}[b]{0.35\linewidth}
				\centering
				\includegraphics[width=\linewidth]{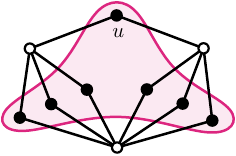}
				\caption{An optimal \textsc{Type2} cluster of size~$7$ that is disconnected and has diameter~$2$. Every connected cluster of this size has diameter~$3$, so the bound of \cref{lem:diameter_increas_conn} is tight.}
				\label{fig:unrestrictedDisconnected}
			\end{subfigure}
			\caption{Two examples for Best-or-Fill. All edges have unit weight.}
			\label{fig:best-or-fill}
		\end{figure}

		The best cluster $B_u$ of a node $u$ comes in four types. In each of them, ties are broken arbitrarily, but connected clusters are preferred. A \textsc{Type1} cluster consists of $u$ and the $r-1$ nodes nearest to~$u$. A \textsc{Type2} cluster has size~$r$, contains~$u$, and has the smallest possible diameter. A \textsc{Type3} cluster is obtained from a \textsc{Type2} cluster by connecting its components via shortest paths from~$u$. A \textsc{Type4} cluster has size~$r$, contains~$u$, induces a connected subgraph, and has the smallest possible diameter among such sets. \textsc{Type1}, \textsc{Type3}, and \textsc{Type4} clusters are connected, whereas \textsc{Type2} clusters need not be (\cref{fig:unrestrictedDisconnected}). Computing a minimum-diameter $r$-set is at least as hard as detecting an $r$-clique, since in an unweighted graph an $r$-set of diameter~$1$ is exactly an $r$-clique.
		
		\begin{lemma}\label{lem:diameter_increas_conn}
			The diameter of a \textsc{Type3} cluster is at most $\frac{3}{2}$ times the diameter of its underlying \textsc{Type2} cluster, and this is tight (\cref{fig:unrestrictedDisconnected}).
		\end{lemma}
		
		\begin{proof}[Proof of \cref{thm:approx:Type2cluster}, \textsc{Type2} clusters]
			The approximation guarantee follows as in the proof of \cref{thm:approx:Type1cluster}. The only difference is that we no longer need the detour through the center of a cluster, since a \textsc{Type2} cluster has diameter at most~$d^*$ itself.
			On general graphs, \textsc{Type2} clusters need not be connected.
			On a tree, however, some minimum-diameter $r$-set through $u$ is connected. To see this, take a minimum-diameter $r$-set through $u$ whose minimal spanning subtree has the fewest edges. Any vertex of that subtree outside the set could replace a leaf of the set without increasing the diameter, so the set is the vertex set of a subtree.
		\end{proof}
		
		\begin{proof}[Proof of \cref{lem:diameter_increas_conn}]
			Let $B_u$ be the underlying \textsc{Type2} cluster and let $d$ be its diameter.
			Consider any two nodes $x$ and $y$ of the \textsc{Type3} cluster.
			Since we only used shortest paths to $u$ to connect the components of $B_u$, we can choose two nodes $x'$ and $y'$ from $B_u$ such that $x$ lies on a shortest $x'$-$u$-path and $y$ lies on a shortest $y'$-$u$-path within the (larger) \textsc{Type3} cluster ($x=x'$ or $y=y'$ is possible).
			Let $d_x$, $d_y$ denote distances from $u$ to $x$ and $y$, let $d'_x$ denote the distance between $x$ and $x'$, and let $d'_y$ denote the distance between $y$ and $y'$.
			Then $d_x+d'_x+d_y+d'_y \leq 2d$.
			Observe that $x$-$y$-paths may run via $u$ or via $x'$ and $y'$.
			So the distance $d'$ between $x$ and $y$ is at most $\min\{d_x+d_y, d'_x+d'_y+d\}$.
			Combining these inequalities gives $d' \leq \frac{3}{2}d$.
		\end{proof}
		
		\begin{proof}[Proof of \cref{thm:approx:Type2cluster}, \textsc{Type3} clusters]
			The approximation guarantee follows as in the proofs above.
			Marked clusters have diameter at most $\frac{3}{2}d^*$, and each node outside the marked clusters has distance at most $d^*$ to some marked cluster. (Note that every node of a \textsc{Type3} cluster still has distance at most $d^*$ from~$u$.)
		\end{proof}
		
		\begin{proof}[Proof of \cref{thm:approx:Type2cluster}, \textsc{Type4} clusters]
			Connectedness is as in \cref{thm:approx:Type1cluster}. For the bound, fix $u$ and let $C\ni u$ be its cluster in an optimal connected solution. Restricting a spanning tree of $C$ rooted at $u$ to the $r$ nodes closest to $u$ gives a connected $r$-set through $u$ of diameter at most $d^*_\mathrm{conn}$, so $B_u$ has diameter at most $d^*_\mathrm{conn}$, and so does every marked cluster. Every filled node $v$ lies within $d^*_\mathrm{conn}$ of its marked cluster, since otherwise no node of $B_v$ would lie in an earlier marked cluster and $B_v$ would itself have been marked. Two nodes of the same extended cluster are therefore at distance at most $d^*_\mathrm{conn}+\mathrm{diam}(B)+d^*_\mathrm{conn}\le 3d^*_\mathrm{conn}$.
		\end{proof}
		
		\begin{proof}[Proof of \cref{cor:mst_spanner}]
			The clustering can be constructed with the exact algorithm on trees from \cref{thm:conv_tree_fast}. The guarantees follow from the correctness of the algorithm combined with \cref{thm:spanner} for $\alpha=1$, $\beta=3$ (the latter implied by \cref{lem:separations}\,(ii)).
		\end{proof}
		
				\subsection{The \texorpdfstring{$t$}{t}-Spanner Transfer}\label{sec:app:spanner}
		
		\begin{proof}[Proof of \cref{thm:spanner}]
			We start with $t\alpha$-ratio-convexity.
			When we move from $G_t$ to~$G$, edges are only added, so every path inside a cluster $C$ of $S$ still exists.
			Distances in $G$ may be shorter than in~$G_t$, but since $G_t$ is a $t$-spanner of~$G$, every distance in $G_t$ is at most $t$ times the corresponding distance in~$G$.
			Hence $\mathrm{d}_{G[C]}(v,w)\le\mathrm{d}_{G_t[C]}(v,w)\le\alpha\,\mathrm{d}_{G_t}(v,w)\le t\alpha\,\mathrm{d}_G(v,w)$ for all $v,w\in C$, so $C$ is $t\alpha$-ratio-convex in~$G$.
			Moreover, $S$ is an $r$-gather of~$G$, since its clusters are unchanged.
			Finally, we prove the approximation bound.
			Let $S^*$ denote any solution of (unrestricted) $r$-gather in $G$.
			Then $S^*$ is a solution to (unrestricted) $r$-gather in $G_t$.
			Let $D_S$, $D_{S^*}$ denote the diameter of $S$ and $S^*$ in $G$ and let $D_S^t$, $D_{S^*}^t$ denote the diameter of $S$ and $S^*$ in $G_t$.
			Then $D_S \leq D^t_S \leq \beta D^t_{S^*} \leq t\beta D_{S^*}$.
		\end{proof}

		\section{Deferred Material for Section~\ref{sec:treewidth}}\label{sec:app:tw}

		This appendix gives the construction behind \cref{thm:tw-dp,cor:tw-fpt}.
		
		\subsection{Separators and Distance Profiles}\label{sec:tw-profiles}
		
		We use nice tree decompositions in the standard form~\cite{cygan2015parameterized}. Such a decomposition is a rooted tree $T$ with bags $X_t\subseteq V$, of width $\mathrm{tw}=\max_t\lvert X_t\rvert-1$, with empty bags at the leaves and at the root, and with every non-leaf node being an \emph{introduce} node ($X_t=X_{t'}\cup\{v\}$, $v\notin X_{t'}$), a \emph{forget} node ($X_t=X_{t'}\setminus\{v\}$, $v\in X_{t'}$) or a \emph{join} node ($X_t=X_{t_1}=X_{t_2}$ and $V_{t_1}\cap V_{t_2}=X_t$), where $V_t$ denotes the union of the bags in the subtree rooted at~$t$. Such a decomposition with $O(\mathrm{tw}\cdot n)$ nodes can be computed from any tree decomposition in linear time, and one of width at most $2\mathrm{tw}+1$ can be computed in $2^{O(\mathrm{tw})}n$ time~\cite{korhonen2021single}. We assume that such a decomposition is given. Besides the node types, we use only that $X_t$ separates $V_t\setminus X_t$ from $V\setminus V_t$ at every node~$t$, that $v\notin V_{t'}$ at an introduce node (which holds in every tree decomposition), and that $V_{t_1}\cap V_{t_2}=X_t$ at a join node. In particular, no edge joins $V_t\setminus X_t$ and $V\setminus V_t$.
		
		In a general graph, the following identity takes the role of the decomposition $\mathrm{d}_T(x,y)=\mathrm{d}_T(x,u)+\mathrm{d}_T(u,y)$ of a tree along a cut vertex~$u$, which the dynamic program of \cref{sec:poly_trees} uses implicitly.
		
		\begin{lemma}\label{lem:tw-port}
			Let $t$ be a node of the decomposition, $x\in V_t$ and $y\in V\setminus V_t$. Then
			\[\mathrm{d}_G(x,y)=\min_{b\in X_t}\big(\mathrm{d}_G(x,b)+\mathrm{d}_G(b,y)\big).\]
		\end{lemma}
		\begin{proof}
			The inequality $\le$ follows from the triangle inequality. For $\ge$, a shortest $x$-$y$-path meets $X_t$ because $X_t$ separates $V_t\setminus X_t$ from $V\setminus V_t$, and splitting it at such a vertex gives the bound. (If no $x$-$y$-path exists, both sides are~$\infty$.) This is the identity underlying the center-types of~\cite{blavzej2023parameterized}, where it plays the same role for a cluster center.
		\end{proof}
		
		Both terms on the right-hand side are distances in the whole graph~$G$, not distances inside the processed part. This makes the identity usable, and it is the reason why the state below records distances from $x$ to the bag rather than distances realized inside $G_t$.
		
		\begin{definition}\label{def:tw-profile}
			Let $L_d=\{0,1,\dots,d\}\cup\{\infty\}$, ordered by $0<1<\dots<d<\infty$ and with $a+\infty=\infty$. For a node $t$ and a vertex $x\in V$, the \emph{profile} of $x$ at $t$ is the vector $p_t(x)\in L_d^{X_t}$ with
			\[p_t(x)_b=\begin{cases}\mathrm{d}_G(x,b) & \text{if } \mathrm{d}_G(x,b)\le d,\\ \infty &\text{otherwise.}\end{cases}\]
			For $p,q\in L_d^{X_t}$ we write $m(p,q)=\min_{b\in X_t}(p_b+q_b)\in\mathbb{N}\cup\{\infty\}$ and order $L_d^{X_t}$ componentwise.
		\end{definition}
		
		\begin{lemma}\label{lem:tw-profiledecides}
			Let $t$ be a node, $x\in V_t$ and $y\in V\setminus V_t$. Then
			\begin{enumerate}
				\item[(i)] $\mathrm{d}_G(x,y)\le d$ if and only if $m(p_t(x),p_t(y))\le d$, and then the two are equal.
				\item[(ii)] $m$ is monotone in each argument: $p\le p'$ implies $m(p,q)\le m(p',q)$.
			\end{enumerate}
		\end{lemma}
		\begin{proof}
			Part (ii) is immediate. For (i), if $\mathrm{d}_G(x,y)\le d$ and $b$ realizes the minimum of \cref{lem:tw-port}, then $\mathrm{d}_G(x,b)$ and $\mathrm{d}_G(b,y)$ are at most $d$, so neither entry is truncated and $m$ equals $\mathrm{d}_G(x,y)$. Conversely no port can beat $\mathrm{d}_G(x,y)$, since truncation only raises entries to $\infty$. If $\mathrm{d}_G(x,y)>d$, every port sums to more than~$d$ by \cref{lem:tw-port}, whether or not an entry is truncated.
		\end{proof}
		
		The truncation bounds the alphabet. Since $\lvert L_d\rvert=d+2$, a profile is one of $(d+2)^{\mathrm{tw}+1}$ vectors, independently of~$n$. Moreover, the distances from a fixed $b$ take at most $n$ distinct values, so each coordinate takes at most $\min(d+2,n+1)$ values. From now on, we write $\lambda=\min(d+2,n+1)$.
		
		This bound relies on integral edge lengths, which ensure that all distances up to~$d$ lie in $\{0,\dots,d\}$. For real edge lengths, one would instead keep the $\Theta(n^2)$ realized distances. The state bound of \cref{thm:tw-dp} would then read $(n^2)^{O(\mathrm{tw})}$, and $d$ would no longer be a meaningful parameter, since scaling all lengths rescales it.

		\subsection{Partial Solutions and the State}\label{sec:tw-state}
		
		\begin{definition}\label{def:tw-partial}
			A \emph{partial solution} at a node $t$ is a partition $\mathcal{P}$ of $V_t$ such that every part $Q\in\mathcal{P}$ has $\mathrm{diam}_G(Q)\le d$ and, in addition,
			\begin{itemize}
				\item if $Q\cap X_t=\emptyset$ (a \emph{finished} part), then $G[Q]$ is connected and $\mathrm{W}(Q)\ge r$;
				\item if $Q\cap X_t\neq\emptyset$ (an \emph{open} part), then every connected component of $G[Q]$ contains a vertex of $X_t$.
			\end{itemize}
			A $d$-valid $r$-gather $\mathcal{S}$ of $G$ \emph{extends} $\mathcal{P}$ if $\mathcal{P}=\{C\cap V_t : C\in\mathcal{S},\ C\cap V_t\neq\emptyset\}$, and $\mathcal{P}$ is \emph{extendable} if some $d$-valid $r$-gather extends it.
		\end{definition}
		
		\begin{lemma}\label{lem:tw-restriction}
			If $\mathcal{S}$ is a $d$-valid $r$-gather of $G$, then $\mathcal{P}=\{C\cap V_t: C\in\mathcal{S},\ C\cap V_t\neq\emptyset\}$ is a partial solution at~$t$.
		\end{lemma}
		\begin{proof}
			Each part inherits $\mathrm{diam}_G\le d$. If $Q=C\cap V_t$ misses $X_t$, then $C\subseteq V_t$, since a vertex of $C$ outside $V_t$ would force an edge of the connected graph $G[C]$ between $V_t\setminus X_t$ and $V\setminus V_t$. So $Q=C$ is a finished cluster. If $Q$ meets $X_t$ and some component $D$ of $G[Q]$ does not, then no edge of $G[C]$ leaves $D$, so $D=C=Q$, contradicting $Q\cap X_t\neq\emptyset$.
		\end{proof}
		
		\begin{definition}\label{def:tw-statedef}
			Let $\mathcal{P}$ be a partial solution at $t$. For an open part $Q$, put
			\begin{itemize}
				\item $\beta(Q)=Q\cap X_t$, the \emph{class};
				\item $\gamma(Q)$, the partition of $\beta(Q)$ into the traces of the connected components of $G[Q]$;
				\item $\Pi(Q)$, the set of componentwise maximal elements of $\{p_t(x): x\in Q\}\subseteq L_d^{X_t}$;
				\item $\omega(Q)=\min(\mathrm{W}(Q\setminus X_t),\,r)$, the \emph{saturated private weight}.
			\end{itemize}
			The \emph{state} $\sigma_t(\mathcal{P})$ is the set of quadruples $(\beta(Q),\gamma(Q),\Pi(Q),\omega(Q))$ over the open parts~$Q$. The finished parts contribute nothing.
		\end{definition}
		
		Two conventions in \cref{def:tw-statedef} deserve a comment. First, $\Pi(Q)$ ranges over \emph{all} vertices of~$Q$, including those outside the bag. This is necessary because the open diameter constraints concern exactly the processed vertices outside the bag, so distances between bag vertices alone do not suffice. Second, we record the weight of $Q\setminus X_t$, which is the part of $Q$ that no later step counts again. At a join node, $Q_1$ and $Q_2$ share exactly $\beta(Q)$. Hence their private weights add up, and the saturation of \cref{lem:dp-saturation} applies to the sum. Saturated total weights could not be combined in this way, since $\min(\cdot,r)$ does not commute with subtraction.
		
		\begin{theorem}\label{thm:tw-state}
			Let $\mathcal{P}$ and $\mathcal{P}'$ be partial solutions at a node $t$ with $\sigma_t(\mathcal{P})=\sigma_t(\mathcal{P}')$. Then $\mathcal{P}$ is extendable if and only if $\mathcal{P}'$ is extendable.
		\end{theorem}
		\begin{proof}
			By symmetry, let $\mathcal{S}$ be a $d$-valid $r$-gather extending $\mathcal{P}$. The classes $\beta(Q)$ of the open parts of $\mathcal{P}$ are non-empty and partition $X_t$, and likewise for $\mathcal{P}'$. Distinct open parts therefore have distinct classes and hence distinct quadruples, so the two states have exactly as many elements as there are open parts and the matching is forced. There is a bijection $\varphi$ from the open parts of $\mathcal{P}$ to those of $\mathcal{P}'$ with $\beta(\varphi(Q))=\beta(Q)$, $\gamma(\varphi(Q))=\gamma(Q)$, $\Pi(\varphi(Q))=\Pi(Q)$ and $\omega(\varphi(Q))=\omega(Q)$. Write $\mathcal{F}'$ for the set of finished parts of $\mathcal{P}'$ and put
			\[\mathcal{S}'=\{C\in\mathcal{S}: C\cap V_t=\emptyset\}\ \cup\ \big\{(C\setminus V_t)\cup\varphi(C\cap V_t) : C\in\mathcal{S},\ C\cap V_t \text{ open}\big\}\ \cup\ \mathcal{F}'.\]
			Every $C\in\mathcal{S}$ with $C\cap V_t\neq\emptyset$ meets $V_t$ in a part of $\mathcal{P}$, by the definition of \enquote{extends}, and $C\subseteq V_t$ holds for every $C$ whose trace is finished, by \cref{lem:tw-restriction}. So no cluster with a finished trace contributes a vertex outside~$V_t$, and the three blocks of $\mathcal{S}'$ are disjoint and cover~$V$. (A cluster with an \emph{open} trace may also satisfy $C\subseteq V_t$. Then $C\setminus V_t=\emptyset$ and $C'=\varphi(Q)$, which the arguments below cover without change.) Hence $\mathcal{S}'$ is a partition of $V$ that extends $\mathcal{P}'$, and it remains to check that each of its parts is a cluster of a $d$-valid $r$-gather. The parts of $\mathcal{F}'$ are connected, of weight at least $r$ and of diameter at most $d$ by \cref{def:tw-partial}, and the clusters disjoint from $V_t$ are unchanged. So let $C\in\mathcal{S}$ with $Q=C\cap V_t$ open and let $C'=(C\setminus V_t)\cup\varphi(Q)$.
			\emph{Weight.} Put $a=\mathrm{W}(C\setminus V_t)+\mathrm{W}(\beta(Q))\ge 0$ and $b=\mathrm{W}(\varphi(Q)\setminus X_t)\ge0$. Using $\min(a+b,r)=\min(a+\min(b,r),r)$ for $a,b\ge0$, the identity of \cref{lem:dp-saturation}, and $\beta(\varphi(Q))=\beta(Q)$, we get
			\[\min(\mathrm{W}(C'),r)=\min\big(a+\omega(\varphi(Q)),r\big)=\min\big(a+\omega(Q),r\big)=\min(\mathrm{W}(C),r)=r,\]
			the last equality because $\mathcal{S}$ is an $r$-gather. Hence $\mathrm{W}(C')\ge r$.
			\emph{Connectivity.} No edge joins $V_t\setminus X_t$ and $V\setminus V_t$, so in $G[C']$ every edge between $C\setminus V_t$ and $\varphi(Q)$ has its endpoint in $\beta(Q)$. The same holds for $G[C]$ with $Q$ in place of $\varphi(Q)$. Moreover, the components of $G[\varphi(Q)]$ correspond bijectively to the classes of $\gamma(\varphi(Q))=\gamma(Q)$. Every component meets $X_t$ by the open-part condition of \cref{def:tw-partial}, so every trace is non-empty and the trace map is injective. It is surjective because $\gamma$ is defined as the partition into these traces. The same is true for $G[Q]$. This is the step for which the open-part condition is indispensable. Consider the auxiliary graph $H$ whose vertices are the components of $G[C\setminus V_t]$ together with the classes of $\gamma(Q)$, with an edge between a component $D$ and a class $g$ whenever some vertex of $D$ has a neighbor in~$g$. Contracting each component of $G[C\setminus V_t]$ and each component of $G[Q]$ turns $G[C]$ into $H$ without losing any connection. Edges inside a component disappear, there are no edges between two components on the same side, and every edge between the two sides has an endpoint in $\beta(Q)$, as shown above. Hence $G[C]$ is connected if and only if $H$ is connected, and the same equivalence holds for $G[C']$ with the same graph~$H$, since $H$ depends only on $C\setminus V_t$, on $\beta(Q)$ and on $\gamma(Q)$, and $G[X_t]$ is the same graph in both cases. As $G[C]$ is connected, so is $G[C']$.
			\emph{Diameter.} Pairs inside $C\setminus V_t$ have both endpoints in $C$ and are therefore bounded by $\mathrm{diam}_G(C)\le d$, and pairs inside $\varphi(Q)$ satisfy $\mathrm{d}_G\le d$ because $\mathcal{P}'$ is a partial solution. Let $x\in\varphi(Q)$ and $y\in C\setminus V_t$, so $x\in V_t$ and $y\notin V_t$. Choose $p\in\Pi(\varphi(Q))$ with $p_t(x)\le p$. Since $\Pi(\varphi(Q))=\Pi(Q)$, there is $x_0\in Q$ with $p_t(x_0)=p$. As $x_0,y\in C$ we have $\mathrm{d}_G(x_0,y)\le d$, so $m(p,p_t(y))\le d$ by \cref{lem:tw-profiledecides}(i), and therefore $m(p_t(x),p_t(y))\le m(p,p_t(y))\le d$ by monotonicity (ii). Applying \cref{lem:tw-profiledecides}(i) once more gives $\mathrm{d}_G(x,y)\le d$. Hence $\mathrm{diam}_G(C')\le d$.
		\end{proof}
		
		The weight coordinate of the state is \emph{saturated} at~$r$. On trees, this corresponds to the following observation about the table entries of \cref{sec:poly_trees}. It does not improve the bounds of \cref{thm:conv_tree_fast}, but we state it here because it generalizes from cut vertices to separators.
		
		\begin{lemma}\label{lem:dp-saturation}
			Replacing every table entry $f_u(\delta)$ by $\min(f_u(\delta),r)$, with $-\infty$ absorbing, does not change the outcome. For integral $r$ and integral node weights, the saturated $f_u$ then takes at most $r+2$ distinct values, independently of $\lvert T_u\rvert$ and of the edge weights.
		\end{lemma}
		\begin{proof}
			An entry is summed into an ancestor's root-cluster weight, maximized, tested against $-\infty$, and compared with~$r$. All four operations are compatible with saturation, since $\min(a+b,r)=\min(a+\min(b,r),r)$ for $a,b\ge0$, $\min(\cdot,r)$ commutes with $\max$, $\min(-\infty,r)=-\infty$, and $x\ge r\iff\min(x,r)\ge r$. Saturation applies to entries only, never to the partial sum $S$ above, since subtraction does not commute with $\min(\cdot,r)$. For integral weights, the saturated entries therefore lie in $\{-\infty,0,1,\dots,r\}$.
		\end{proof}
		
		\begin{corollary}\label{cor:tw-weight}
			Let $\mathcal{P},\mathcal{P}'$ be partial solutions at $t$ and $\varphi$ a bijection between their open parts with $\beta,\gamma,\Pi$ equal classwise. Then:
			\begin{enumerate}
				\item[(i)] if $\omega(\varphi(Q))=\omega(Q)$ for all $Q$, then $\mathcal{P}$ is extendable if and only if $\mathcal{P}'$ is;
				\item[(ii)] if $\omega(\varphi(Q))\ge\omega(Q)$ for all $Q$ and $\mathcal{P}$ is extendable, then so is $\mathcal{P}'$.
			\end{enumerate}
			In particular, one saturated weight per class suffices, instead of a set of Pareto-optimal (profile, weight) pairs. By~(ii), a state may moreover be discarded if another state with the same $\beta,\gamma,\Pi$ has a componentwise larger $\omega$-vector; this pruning is optional and not needed for the bound. This coordinate contributes $(r+1)^{\mathrm{tw}+1}$ instead of $(\mathrm{W}(V)+1)^{\mathrm{tw}+1}$ possibilities.
		\end{corollary}
		\begin{proof}
			Part (i) is \cref{thm:tw-state}, since $\omega$ is the only coordinate of the state in which the weights enter. Part~(ii) follows from the proof of \cref{thm:tw-state} with one inequality relaxed. In the weight paragraph, we now have $\min(\mathrm{W}(C'),r)=\min(a+\omega(\varphi(Q)),r)\ge\min(a+\omega(Q),r)=r$, and no other paragraph of that proof uses~$\omega$. For the counting, note that $\omega(Q)\in\{0,1,\dots,r\}$ for integral weights and that there are at most $\lvert X_t\rvert\le\mathrm{tw}+1$ open parts.
		\end{proof}
		
		\Cref{cor:tw-weight} is the separator analog of \cref{lem:dp-saturation}. It shows that the size constraint does \emph{not} force a factor of $n^{O(\mathrm{tw})}$. Whether the distance coordinate can be compressed in a similar way is open (\cref{sec:treewidth}).

		\subsection{The Dynamic Program}\label{sec:tw-dp}
		
		For every node $t$, let $A_t$ be the set of states of partial solutions at~$t$. The instance is a yes-instance if and only if the empty state belongs to~$A_{\mathrm{root}}$. Indeed, since $V_{\mathrm{root}}=V$ and $X_{\mathrm{root}}=\emptyset$, a partial solution at the root is precisely a $d$-valid $r$-gather. All of its parts are finished, so its state is empty. The correctness of the algorithm relies only on \cref{lem:tw-transitions}. \Cref{thm:tw-state} justifies the choice of the state and is the basis of \cref{cor:tw-weight}, but the algorithm does not use it. We compute the sets $A_t$ bottom-up and manipulate each state class by class. For a set $\mathcal{V}\subseteq L_d^{X}$, we write $\max\mathcal{V}$ for the set of its componentwise maximal elements.
		
		A leaf node has $X_t=V_t=\emptyset$ and $A_t=\{\emptyset\}$.
		
		\paragraph{Introduce node ($X_t=X_{t'}\cup\{v\}$).} Every stored profile gains a coordinate for~$v$, which is computed by \cref{lem:tw-port}. For $p\in L_d^{X_{t'}}$, let $\mathrm{ext}(p)\in L_d^{X_t}$ agree with $p$ on $X_{t'}$ and satisfy $\mathrm{ext}(p)_v=m(p,p_{t'}(v))$ if this value is at most~$d$ and $\infty$ otherwise. From each state in $A_{t'}$ we produce states of $A_t$ as follows. First decide whether $v$ opens a new class or joins an existing class~$c$. The latter is allowed only if $m(p,p_{t'}(v))\le d$ for all $p\in\Pi(c)$, a test performed on the profiles as stored, that is, over~$X_{t'}$. Then set $\Pi(c)\mathrel{{:}{=}}\max(\{\mathrm{ext}(p):p\in\Pi(c)\}\cup\{p_t(v)\})$ for the class $c$ of~$v$ and $\Pi(c'')\mathrel{{:}{=}}\max\{\mathrm{ext}(p):p\in\Pi(c'')\}$ for every other class $c''$, so that all profiles are indexed by $X_t$ afterwards. A new class $\{v\}$ receives $\beta=\{v\}$, $\gamma=\{\{v\}\}$, $\Pi=\{p_t(v)\}$ and $\omega=0$. Joining an existing class $c$ sets $\beta(c)\mathrel{{:}{=}}\beta(c)\cup\{v\}$, leaves $\omega(c)$ unchanged, and merges into one group of $\gamma(c)$ the singleton $\{v\}$ together with all groups containing a neighbor of~$v$.
		
		\paragraph{Forget node ($X_t=X_{t'}\setminus\{v\}$).} Let $c$ be the class of $v$ and let $g$ be its group in $\gamma(c)$. First re-index all profiles, in \emph{every} class and before any of the branches below, by $\Pi(c'')\mathrel{{:}{=}}\max\{p|_{X_t}: p\in\Pi(c'')\}$, and set $\omega(c)\mathrel{{:}{=}}\min(\omega(c)+\mathrm{w}_v,r)$, since $v$ becomes private. If $\beta(c)=\{v\}$, the class is \emph{closed}. In this case, we discard the state unless $\omega(c)=r$, and otherwise we delete the class from the state. If $\beta(c)\neq\{v\}$ and $g=\{v\}$, we discard the state. Otherwise, we delete $v$ from $\beta(c)$ and from~$g$.
		
		\paragraph{Join node ($X_t=X_{t_1}=X_{t_2}$).} Combine a state of $A_{t_1}$ with a state of $A_{t_2}$ whenever they induce the same classes on $X_t$. For each class~$c$ set $\Pi(c)\mathrel{{:}{=}}\max(\Pi_1(c)\cup\Pi_2(c))$, let $\gamma(c)$ be the finest partition of $\beta(c)$ coarsening both $\gamma_1(c)$ and $\gamma_2(c)$, and set $\omega(c)\mathrel{{:}{=}}\min(\omega_1(c)+\omega_2(c),r)$. Discard the combination unless $m(p_1,p_2)\le d$ for all classes $c$ and all $p_1\in\Pi_1(c)$, $p_2\in\Pi_2(c)$. \Cref{tab:tw-rules} summarizes the four rules.

		\begin{table}[t]
			\centering\small
			\caption{Transitions of the treewidth DP for one state. Here $\max$ keeps the componentwise maximal profiles of a set.}\label{tab:tw-rules}
			\begin{tabular}{@{}>{\raggedright\arraybackslash}p{0.13\linewidth}>{\raggedright\arraybackslash}p{0.56\linewidth}>{\raggedright\arraybackslash}p{0.26\linewidth}@{}}
				\hline
				node & update & discard if\\\hline
				leaf & $A_t=\{\emptyset\}$ & --\\
				introduce $v$ & $v$ opens the class $(\{v\},\{\{v\}\},\{p_t(v)\},0)$, or $v$ joins a class~$c$: $\beta(c)\gets\beta(c)\cup\{v\}$, $v$ merges with all groups of $\gamma(c)$ containing a neighbor, and $\Pi(c)\gets\max(\mathrm{ext}(\Pi(c))\cup\{p_t(v)\})$. Every other class: $\Pi\gets\max\mathrm{ext}(\Pi)$
				& $v$ joins $c$ although $m(p,p_{t'}(v))>d$ for some $p\in\Pi(c)$\\
				forget $v$ & all classes: $\Pi\gets\max(\Pi|_{X_t})$. Class $c\ni v$: $\omega(c)\gets\min(\omega(c)+\mathrm{w}_v,r)$, remove $v$ from $\beta(c)$ and from its group; delete $c$ if $\beta(c)=\{v\}$
				& $\beta(c)=\{v\}$ and $\omega(c)<r$; or $\beta(c)\ne\{v\}$ and the group of $v$ is $\{v\}$\\
				join & pair states with equal classes; $\Pi\gets\max(\Pi_1\cup\Pi_2)$, $\gamma\gets$ finest common coarsening of $\gamma_1,\gamma_2$, $\omega\gets\min(\omega_1+\omega_2,r)$
				& $m(p_1,p_2)>d$ for some class and some $p_i\in\Pi_i$\\\hline
			\end{tabular}
		\end{table}
		
		\begin{lemma}\label{lem:tw-transitions}
			The rules above compute $A_t$ correctly for every node~$t$.
		\end{lemma}
		\begin{proof}
			The bookkeeping of classes and groups is the same as for \textsc{Steiner Tree}~\cite[Sec.~7.3]{cygan2015parameterized} and for connected microaggregation~\cite{blavzej2023parameterized}. Groups merge at an introduce node when the new vertex has neighbors in several of them, and at a join node along shared bag vertices. A vertex may not be forgotten while it is the last bag vertex of its group, unless its class is closed. We verify the parts that are specific to the diameter objective, namely the profile and weight coordinates. We also check that each rule is well defined on states, which is the congruence property the algorithm needs. The proof does not use \cref{thm:tw-state}.
			Throughout, we use that $\max$ commutes with monotone maps. If $f$ is monotone, then $\max f(\mathcal{V})=\max f(\max\mathcal{V})$, since every element of $\mathcal{V}$ is dominated by a maximal one. Both $\mathrm{ext}$ and the restriction $p\mapsto p|_{X_t}$ are monotone, so the updates of $\Pi$ are correct, and a test of the form \enquote{$m(p,q)\le d$ for all profiles $p$ of $Q$} may be run on $\Pi(Q)$ alone.
			At an introduce node, since $v\notin V_{t'}$, all its neighbors in $V_t$ lie in $X_{t'}$, and $\mathrm{ext}$ computes the new coordinate correctly by \cref{lem:tw-port}. Adding $v$ to a part $Q$ keeps the partial solution valid if and only if $\mathrm{d}_G(x,v)\le d$ for all $x\in Q$, which is the test performed by \cref{lem:tw-profiledecides}(i). The vertex must be added to some part, and it is never added to a finished part. Such a part is connected, misses $X_{t'}$, and contains no neighbor of~$v$, so adding $v$ would create a component that misses the bag. Conversely, deleting $v$ leaves a valid partial solution at $t'$, since every component of $G[Q\setminus\{v\}]$ that was split off contains a neighbor of~$v$ and hence a vertex of $X_{t'}$.
			At a forget node, $V_t=V_{t'}$ and only the conditions change. A part with $Q\cap X_{t'}=\{v\}$ becomes finished, which is admissible exactly if $G[Q]$ is connected and $\mathrm{W}(Q)\ge r$. Connectivity is automatic, as every component met $\{v\}$, and the weight condition is the test $\omega(c)=r$ after the update, since $Q\setminus X_t=Q$. That update is correct because $\min(a+\mathrm{w}_v,r)=\min(\min(a,r)+\mathrm{w}_v,r)$ for $a,\mathrm{w}_v\ge0$.

			\emph{Join.} Write $\mathcal{P}_i=\{Q\cap V_{t_i}\}$. No edge joins $V_{t_1}\setminus X_t$ and $V_{t_2}\setminus X_t$, so private weights add, and the update of $\omega$ is the identity of \cref{lem:dp-saturation}. The only pairs whose distance is not already bounded in $\mathcal{P}_1$ or $\mathcal{P}_2$ are those with $x\in Q\cap V_{t_1}\setminus X_t$ and $y\in Q\cap V_{t_2}\setminus X_t$. For them $y\notin V_{t_1}$, so \cref{lem:tw-profiledecides}(i) applies and the executed test is exactly $\mathrm{diam}_G(Q)\le d$. The test also compares profiles of bag vertices, which lie outside that hypothesis, but it cannot cause a wrong rejection. For $x\in X_t$, the port $b=x$ gives $m(p_t(x),p_t(y))=\mathrm{d}_G(x,y)$, and $x,y$ both lie in $Q\cap V_{t_2}$, whose diameter is at most~$d$.
		\end{proof}

		\begin{proof}[Proof of \cref{thm:tw-dp}]
			Correctness follows from \cref{lem:tw-transitions} together with the observation about the root. The distances are computed once, by $n$ runs of Dijkstra's algorithm in $O(n\lvert E\rvert+n^2\log n)$ time, or by $n$ breadth-first searches if $G$ is unweighted.
			Fix a node $t$ and put $k=\lvert X_t\rvert\le\mathrm{tw}+1$. Counting the class and group partitions by Bell numbers, as in~\cite{blavzej2023parameterized,cygan2015parameterized}, gives $B(k)^2\le 2^{O(k\log k)}$ possibilities for $(\beta,\gamma)$. Each $\Pi(c)$ is an antichain among the at most $\lambda^k$ realized profiles, which leaves $2^{\lambda^k}$ possibilities per class and $2^{k\lambda^k}$ over all classes, and $\lvert\Pi(c)\rvert\le P:=\min(n,\lambda^{k})$, since profiles come from vertices. Each $\omega(c)$ is one of $r+1$ values by \cref{cor:tw-weight}. Hence, using $\mathrm{tw}+1\le\lambda^{\mathrm{tw}+1}$,
			\[\lvert A_t\rvert\le 2^{O(\mathrm{tw}\log \mathrm{tw})}\,2^{(\mathrm{tw}+1)\lambda^{\mathrm{tw}+1}}(r+1)^{\mathrm{tw}+1}=2^{\lambda^{O(\mathrm{tw})}}(r+1)^{O(\mathrm{tw})}=:N.\]
			Processing one state at an introduce or forget node costs $O(\mathrm{tw}^2P+\mathrm{tw}P^2)$. Every entry of every stored profile is updated once, and the maximal elements are extracted by pairwise comparison. This comparison is necessary because restricting profiles can destroy incomparability. A join node processes pairs of states, at $O(\mathrm{tw}^2P^2)$ each. Since squaring only doubles the exponent, $N^2P^2=2^{\lambda^{O(\mathrm{tw})}}(r+1)^{O(\mathrm{tw})}$, and with $O(\mathrm{tw}\cdot n)$ nodes the total is as claimed.
			Finally, feasibility is monotone in~$d$ and $d^*_\mathrm{conn}$ is one of the $O(n^2)$ pairwise distances, including~$0$. If $\mathrm{W}(V)<r$ no $r$-gather exists. Binary search needs only $O(\log n)$ calls, but it may query any candidate up to $\mathrm{diam}_G(V)$, and large values of $d$ are expensive because of the dependence of the running time on~$d$. Scanning the candidates in increasing order instead gives at most $O(n^2)$ calls, each with $d\le d^*_\mathrm{conn}$, the last of which dominates.
		\end{proof}

		\begin{proof}[Proof of \cref{cor:tw-fpt}]
			The first claim is \cref{thm:tw-dp}. For the second claim, \cref{lem:UpperBound} applies because all node weights are~$1$, and it gives $d^*_\mathrm{conn}\le 2r-2$ since $\ell_{\max}=1$. So it suffices to run the decision procedure for $d=1,\dots,2r-2$ in increasing order, which is non-empty because $r>1$, and to stop at the first feasible value. Every call then uses $d\le 2r-2$, and the running time becomes $2^{(2r)^{O(\mathrm{tw})}}n+O(n\lvert E\rvert)$.
		\end{proof}

\fi
	\end{document}